\documentstyle[11pt,aaspp4]{article}

\def\la{\mathrel{\hbox{\rlap{\hbox{\lower4pt\hbox{$\sim$}}}\hbox{$<$}}}}
\def\ga{\mathrel{\hbox{\rlap{\hbox{\lower4pt\hbox{$\sim$}}}\hbox{$>$}}}}
\def\lesssim{\mathrel{\hbox{\rlap{\hbox{\lower4pt\hbox{$\sim$}}}\hbox{$<$}}}}
\def\etal{et al.\,\,}

\slugcomment{Accepted to {\it The Astrophysical Journal}} 

\begin{document}

\title{Gamma-Ray Bursts as a Probe of the Very High Redshift Universe}

\author{Donald Q. Lamb and Daniel E. Reichart}

\affil{Department of Astronomy and Astrophysics, University of Chicago,
5640 South Ellis Avenue, Chicago, IL 60637}

\begin{abstract} There is increasingly strong evidence that gamma-ray
bursts (GRBs) are associated with star-forming galaxies, and occur near
or in the star-forming regions of these galaxies.  These associations
provide indirect evidence that at least the long GRBs detected by
BeppoSAX are a result of the collapse of massive stars.  The recent
evidence that the light curves and the spectra of the afterglows of GRB
970228 and GRB 980326 appear to contain a supernova component, in
addition to a relativistic shock wave component, provide more direct
clues that this is the case.  We show that, if many GRBs are indeed
produced by the collapse of massive stars, GRBs and their afterglows
provide a powerful probe of the very high redshift ($z \gtrsim 5$)
universe.  We first establish that GRBs and their afterglows are both
detectable out to very high redshifts.  We then show that one expects
GRBs to occur out to at least $z \approx 10$ and possibly $z \approx
15-20$, redshifts that are far larger than those expected for the most
distant quasars.  This implies that there are large numbers of GRBs
with peak photon number fluxes below the detection thresholds of BATSE
and HETE-2, and even below the detection threshold of {\it Swift}.  The
mere detection of very high redshift GRBs would give us our first
information about the earliest generations of stars.  We show that GRBs
and their afterglows can be used as beacons to locate core collapse
supernovae at redshifts $z \gg 1$, and to study the properties of these
supernovae.  We describe the expected properties of the absorption-line
systems and the Ly$\alpha$ forest in the spectra of GRB afterglows, and
discuss various strategies for determining the redshifts of very high
redshift GRBs.  We then show how the absorption-line systems and the
Ly$\alpha$ forest visible in the spectra of GRB afterglows can be used
to trace the evolution of metallicity in the universe, and to probe the
large-scale structure of the universe at very high redshifts.  Finally,
we show how measurement of the Ly$\alpha$ break in the spectra of GRB
afterglows can be used to constrain, or possibly measure, the epoch at
which re-ionization of the universe occurred, using the Gunn-Peterson
test. \end{abstract}


\section{Introduction}

The relatively accurate (3\arcmin) gamma-ray burst (GRB) positions
found using BeppoSAX and disseminated within a day or so led to the
remarkable discoveries that GRBs have X-ray (Costa et al. 1997),
optical (Galama et al. 1997) and radio (Frail \& Kulkarni 1997)
afterglows.  The redshift distances of eight GRBs are currently known,
either directly from absorption lines in the spectra of the afterglow,
or indirectly, from emission lines in the spectra of a galaxy that is
coincident with the position of the X-ray and optical afterglow.  These
redshifts span the range $z = 0.43 - 3.42$, and imply that GRBs are 
perhaps the most luminous and energetic events in the universe (see
Table 1).  

The most widely discussed models of the central engine of GRBs involve
a black hole and an accretion disk, formed either through the core
collapse of a massive star (Woosley 1993, 1996; Paczy\'nski 1998, MacFadyen \& Woosley 1999; Wheeler et al. 1999; MacFadyen, Woosley \& Heger 1999) or
the coalescence of a neutron star -- neutron star or neutron star --
black hole binary (Paczy\'nski 1986; Narayan, Paczy\'nski \& Piran
1992; M\'esz\'aros \& Rees 1993).  The former are expected to occur
near or in the star-forming regions of their host galaxies, while most
of the latter are expected to occur primarily outside of the galaxies
in which they were born.

Castander and Lamb (1998) showed that the light from the host galaxy of
GRB 970228, the first burst for which an afterglow was detected, is
very blue.  This implies that the host galaxy is undergoing copious
star formation and suggests an association between GRB sources and
star-forming galaxies.  Subsequent analyses of the color of this galaxy
(Castander \& Lamb 1999; Fruchter et al. 1999a) and other host galaxies
(Kulkarni et al. 1998; Fruchter 1999) have strengthened this
conclusion, as has the detection of [OII] and Ly${\alpha}$ emission
lines from several host galaxies (see, e.g., Metzger et al. 1997a;
Kulkarni et al. 1998; Bloom et al. 1998).  

The positional coincidences between burst afterglows and the bright
blue regions of the host galaxies (Sahu et al. 1997, Kulkarni et al.
1998, Fruchter 1999, Kulkarni et al. 1999, Fruchter et al. 1999a), and
the evidence for extinction by dust of some burst afterglows (see,
e.g., Reichart 1998; Kulkarni et al. 1998; Lamb, Castander \& Reichart
1999), lend further support to the idea that GRBs are associated with
star formation, as is expected if GRBs are due to the collapse of
massive stars.  However, this evidence is indirect.

Recent tantalizing  evidence that the light curves and spectra of the
afterglows of GRB 980326 (Bloom et al. 1999) and GRB 970228 (Reichart
1999a, Galama et al. 1999b) contain a supernova (SN) component, in
addition to a relativistic shock wave component, provide more direct
clues that at least the long, softer, smoother bursts (Lamb, Graziani
\& Smith 1993; Kouveliotou et al. 1993) detected by BeppoSAX are a
result of the collapse of massive stars. 

In this paper, we show that, if many GRBs are indeed produced by the
collapse of massive stars, GRBs and their afterglows provide a powerful
probe of the very high redshift ($z \gtrsim 5$) universe.  In \S 2, we
establish that both GRBs and their afterglows are detectable out to
very high redshifts.  In \S 3, we then show that one expects GRBs to
occur out to $z \approx 10$ and possibly $z \approx 15-20$, redshifts
that are far larger than those  expected for the most distant quasars. 
This implies that there are large numbers of GRBs with peak photon
number fluxes below the detection thresholds of BATSE and HETE-2, and
even below the detection threshold of {\it Swift}.  The mere detection
of very high redshift GRBs would give us our first information about
the earliest generations of stars.  In \S 4, we show  that GRBs and
their afterglows can be used as beacons to locate core collapse
supernovae at redshifts $z \gg 1$, and to study the properties of these
supernovae.  In \S 5, we describe the expected properties of the
absorption-line systems and the Ly$\alpha$ forest in the spectra of GRB
afterglows, and discuss various strategies for determining the
redshifts of very high redshift GRBs.  We then show in \S 6 how the
absorption-line systems and the Ly$\alpha$ forest visible in the
spectra of GRB afterglows can be used to trace the evolution of
metallicity in the universe, and in \S 7 how they can be used to probe
the large-scale structure of the universe at very high redshifts.  
Finally, in \S 8 we show how measurement of the Ly$\alpha$ break in the
spectra of GRB afterglows can be used to constrain, or possibly
measure, the epoch at which re-ionization of the universe occurred,
using the Gunn-Peterson test.  We summarize our conclusions in \S 9.

\section{Detectability of GRBs and Their Afterglows at Very High
Redshifts}

It is now clear that GRBs are detectable out to very high redshifts
(VHRs).  In order to establish this, we calculate the limiting
redshifts detectable by BATSE and HETE-2, and by {\it
Swift}, for the seven GRBs with well-established redshifts and
published peak photon number fluxes.  The peak photon number luminosity
is
\begin{equation}
L_P = \int_{\nu_l}^{\nu_u}\frac{dL_P}{d\nu}d\nu \; ,
\end{equation}
where $\nu_l < \nu < \nu_u$ is the band of observation.  Typically, 
for the Burst and Transient Source Experiment (BATSE) on the {\it
Compton} Gamma-Ray Observatory, $\nu_l = 50$ keV and $\nu_u = 300$ keV.  The
corresponding peak photon number flux $P$ is 
\begin{equation}
P = \int_{\nu_l}^{\nu_u}\frac{dP}{d\nu}d\nu \; .
\end{equation}
Assuming that GRBs have a photon number spectrum of the form $dL_P/d\nu
\propto \nu^{-\alpha}$ and that $L_P$ is independent of z, the observed
peak photon number flux $P$ for a burst occurring at a redshift
$z$ is given by
\begin{equation}
P = \frac{L_P}{4\pi D^2(z)(1+z)^{\alpha}} \; ,
\end{equation}
where
\begin{equation}
D(z) = c\int_0^z(1+z')\left|\frac{dt(z')}{dz'}\right|dz' \; 
\end{equation}
is the comoving distance to the GRB, and
\begin{equation}
\frac{dt(z)}{dz} = -\frac{c}{H_0}\frac{1}{(1+z)\sqrt{\Omega_m(1+z)^3 + 
\Omega_{\Lambda}+ (1 -\Omega_m-\Omega_{\Lambda})(1+z)^2}} \; .
\end{equation}
Throughout this paper we take $\Omega_m + \Omega_{\Lambda} = 1$.  Then 
\begin{equation}
D(z) = \frac{c}{H_0}\int_0^z\frac{dz'}{\sqrt{\Omega_m(1+z')^3 + 
\Omega_{\Lambda}}} \; .
\end{equation}
Taking $\alpha = 1$, which is typical of GRBs (Mallozzi, Pendleton \&
Paciesas 1996),
\begin{equation}
P = \frac{L_P}{4\pi D^2(z)(1+z)} \; ,
\end{equation}
which is coincidentally identical to the form one gets when 
$P$ and $L_P$ are bolometric quantities.

Using these expressions, we have calculated the limiting redshifts
detectable by BATSE and HETE-2, and by {\it Swift}, for the seven GRBs with well-established redshifts and
published peak photon number fluxes.  In doing so, we have used the
peak photon number fluxes given in Table 1, taken a detection threshold of
0.2 ph s$^{-1}$ for BATSE (Meegan et al. 1993) and HETE-2 (Ricker 1998) 
and 0.04 ph s$^{-1}$ for {\it Swift} (Gehrels 1999), and
set $H_0 =  65$ km s$^{-1}$ Mpc$^{-1}$, $\Omega_m = 0.3$, and
$\Omega_{\Lambda} = 0.7$ (other cosmologies give similar results).

Figure 1 displays the results.  This figure shows that BATSE and
HETE-2  would be able to detect four of these GRBs (GRBs 970228,
970508, 980613, and 980703) out to redshifts $2 \lesssim z \lesssim 4$,
and three (GRBs 971214, 990123, and 990510) out to redshifts of $20
\lesssim z \lesssim 30$. {\it Swift} would be able to detect the former
four out to redshifts of $5 \lesssim z \lesssim 15$, and the latter
three out to redshifts in excess of $z \approx 70$, although it is
unlikely that GRBs occur at such extreme redshifts (see \S 3 below). 
Consequently, if GRBs occur at VHRs, BATSE has probably already
detected them, and future missions should detect them as well.

The soft X-ray, optical and infrared afterglows of GRBs are also
detectable out to VHRs.  The effects of distance and redshift tend to
reduce the spectral flux in GRB afterglows in a given frequency band,
but time dilation tends to increase it at a fixed time of observation
after the GRB, since afterglow intensities tend to decrease with time. 
These effects combine to produce little or no decrease in the spectral
energy flux $F_{\nu}$ of GRB afterglows in a given frequency band and
at a fixed time of observation after the GRB with increasing redshift:
\begin{equation}
F_{\nu}(\nu,t) = \frac{L_{\nu}(\nu,t)}{4\pi D^2(z) (1+z)^{1-a+b}},
\end{equation}
where $L_\nu \propto \nu^at^b$ is the intrinsic spectral luminosity of
the GRB afterglow, which we assume applies even at early times, and $D(z)$ is again the comoving distance to the
burst.   Many afterglows fade like $b \approx -4/3$, which implies that
$F_{\nu}(\nu,t) \propto D(z)^{-2} (1+z)^{-5/9}$ in the simplest
afterglow model where $a = 2b/3$ (see, e.g., Wijers, Rees, \&
M\'esz\'aros 1997).  In addition, $D(z)$ increases very slowly with redshift at redshifts greater than a few.  Consequently, there is
little or no decrease in the spectral flux of GRB afterglows with
increasing redshift beyond $z \approx 3$.  

For example, Halpern et al. (1999) find in the case of GRB 980519 that
$a = -1.05\pm0.10$ and $b = -2.05\pm0.04$ so that $1-a+b = 0.00 \pm
0.11$, which implies no decrease in the spectral flux with increasing
redshift, except for the effect of $D(z)$.  In the simplest afterglow
model where $a = 2b/3$, if the afterglow declines more rapidly than $b
\approx 1.7$, the spectral flux actually {\it increases} as one moves
the burst to higher redshifts! 

As another example, we calculate the best-fit spectral flux
distribution of the early afterglow of GRB 970228 from Reichart
(1999a), as observed one day after the burst, transformed to various
redshifts.  The transformation involves (1) dimming the
afterglow,\footnote{Again, we have set $\Omega_m = 0.3$ and
$\Omega_{\Lambda} = 0.7$; other cosmologies yield similar results.} (2)
redshifting its spectrum, (3) time dilating its light curve, and (4)
extincting the spectrum using a model of the Ly$\alpha$ forest.  
For the model of the Ly$\alpha$ forest, we have adopted the best-fit
flux deficit distribution to Sample 4 of Zuo \& Lu (1993) from Reichart
(1999b).  At redshifts in excess of $z = 4.4$, this model is an
extrapolation, but it is consistent with the results of theoretical
calculations of the redshift evolution of Ly$\alpha$ absorbers (see,
e.g., Valageas, Schaeffer \& Silk 1999).  Finally, we have convolved
the transformed spectra with a top hat smearing function of width
$\Delta \nu = 0.2\nu$.  This models these spectra as they would be
sampled photometrically, as opposed to spectroscopically; i.e., this
transforms the model spectra into model spectral flux distributions.

Figure 2 shows the resulting K-band light curves.  For a fixed
band and time of observation, steps (1) and (2) above dim the afterglow
and step (3) brightens it, as discussed above.  Figure 2 shows that in
the case of the early afterglow of GRB 970228, as in the case of GRB
980519, at redshifts greater than a few the three effects nearly cancel
one another out.  Thus the afterglow of a GRB occurring at a redshift
slightly in excess of $z = 10$ would be detectable at K $\approx 16.2$
mag one hour after the burst, and at K $\approx 21.6$ mag one day after
the burst, if its afterglow were similar to that of GRB 970228 (a
relatively faint afterglow).  

Figure 3 shows the resulting spectral flux distribution.  The spectral
flux distribution of the afterglow is cut off by the Ly$\alpha$ forest
at progressively lower frequencies as one moves out in redshift.  Thus 
high redshift ($1 \lesssim z \lesssim 5$) afterglows are characterized
by an optical ``dropout'' (Fruchter 1999), and very high redshift ($z
\gtrsim 5$) afterglows by an infrared ``dropout.''

We also show in Figure 3 the effect of a moderate ($A_V = 1/3$), fixed
amount of extinction at the redshift of the GRB.  However, the amount
of extinction is likely to be very small at large redshifts because of
the rapid decrease in metallicity beyond $z = 3$ (see \S6 below).

So far, optical observations have been favored over near-infrared (NIR)
and IR observations in afterglow searches.  This is understandable,
given the greater availability of optical cameras and the modest
(typically 2\arcmin $\times$ 2\arcmin) fields-of-view of current NIR
cameras.  Usually, deep NIR observations have been carried out only
once an optical transient has been identified in a GRB error circle,
thereby assuring that the afterglow can be captured within the
field-of-view of the NIR camera.  The K-band afterglow search of
Gorosabel et al. (1998), which detected the afterglow of GRB 971214
only 3.2 hours after the burst, is a notable exception.  Unfortunately,
the current search strategy of waiting for the identification of an
afterglow candidate at optical wavelengths before carrying out NIR
observations biases against the identification of VHR GRBs, since the
afterglows of these bursts cannot be detected at optical wavelengths.  

The results of our calculations show that the identification of VHR
GRBs will require afterglow searches that incorporate on a consistent
basis (1) sufficiently deep NIR observations, carried out within hours
to days after the burst, and (2) near-simultaneous optical observations
that go sufficiently deep to meaningfully constrain the redshift of the
burst, in the event that its afterglow is only detected at NIR
wavelengths.  Fortunately, early NIR observations will be facilitated
by the HETE-2 (Ricker 1998) and {\it Swift} missions
(Gehrels 1999), which will provide positions accurate to better than
several arcminutes for many GRBs in near-real time.

For more than half of the nearly two dozen GRBs for which X-ray
afterglows have been detected, a corresponding optical afterglow has not been
detected.  A possible explanation of this ``missing optical afterglow''
problem is that, because of larger positional error circles or other
reasons, optical afterglow searches do not always go deep enough, soon
enough, to detect the fading afterglows.  This may explain many of the 
missing optical afterglows, but it probably does not account for all of them. 
Another possible explanation is that some of these afterglows are
significantly extincted by dust, either in our galaxy, in the host
galaxies, or in the environments immediate to the bursts themselves
(see, e.g., Reichart 1998, 1999b).  Finally, it is possible that some
of the GRBs for which no optical afterglow has been detected occurred
at VHRs, and therefore their afterglow spectra were absorbed in the
optical, as described above.  In reality, a combination of these three
effects may be at work.  Early HST and NIR afterglow searches,
facilitated by the more accurate near-real time positions that the HETE-2
and {\it Swift} missions will provide, could help to
distinguish between these various explanations.

In conclusion, if GRBs occur at very high redshifts, both they and
their afterglows would be detectable.

\section{GRB Afterglows as a Probe of Star Formation}

The positional coincidences between burst afterglows and the bright
blue regions of the host galaxies (Sahu et al. 1997, Kulkarni et al.
1998, Fruchter 1999, Kulkarni et al. 1999, Fruchter et al. 1999a), and
the evidence for extinction by dust of some burst afterglows (see,
e.g., Reichart 1998; Kulkarni et al. 1998; Lamb, Castander \& Reichart
1999), lends support to the idea that GRBs are associated with star
formation.  

The discovery of what appears to be supernova components in the
afterglows of GRBs 970228 (Reichart
1999a; Galama et al. 1999b) and 980326 (Bloom et al. 1999) strongly suggests that at least some GRBs
are related to the deaths of massive stars, as predicted by the
widely-discussed collapsar model of GRBs (see, e.g., Woosley 1993,
1996; Paczy\'nski 1998; MacFadyen \& Woosley 1999; Wheeler et al. 1999; MacFadyen, Woosley \& Heger 1999).  The presence of an unusual radio supernova,
SN 1998bw, in the error circle of GRB 980425 (Galama et al. 1998;
Kulkarni et al. 1998) also lends support to this hypothesis, although
the identification of SN 1998bw with GRB 980425 is not secure (see,
e.g., Graziani, Lamb, \& Marion 1999).  If GRBs are related to the
collapse of massive stars, one expects the GRB rate to be approximately
proportional to the star formation rate (SFR).

Observational estimates (Gallego et al. 1995; Lilly et al. 1996;
Connolly et al. 1997; Madau, Pozzetti \& Dickinson 1998) indicate that
the SFR in the universe was about 15 times larger
at a redshift $z \approx 1$ than it is today.  The data at higher
redshifts from the Hubble Deep Field (HDF) in the north suggests a peak in the SFR
at $z \approx 1-2$ (Madau, Pozzetti \& Dickinson 1998), but the actual
situation is highly uncertain.  Assuming that GRBs are standard candles
and the estimate of the SFR derived by Madau, Pozzetti \& Dickinson
(1998; however, see Pei, Fall \& Hauser 1999), several authors (Totani 1997, 1999; Wijers et al. 1998) have
investigated whether or not the observed GRB brightness distribution is
consistent with such a SFR, which rises rapidly from the present epoch
and peaks at $z \approx 1-2$.  Totani (1997, 1999) finds that it is not, and one can infer from the results of Wijers et al. (1998) that it is not.

However, there is now overwhelming evidence that GRBs are not standard
candles:  The peak photon number fluxes $P$ of the seven GRBs with
secure redshifts and published peak photon fluxes span nearly two
orders of magnitude (see Table 1).  The range of peak photon number
fluxes may actually be much greater (Loredo \& Wasserman 1998). 
Furthermore, theoretical calculations show that the birth rate of Pop
III stars produces a peak in the star-formation rate in the universe at
redshifts $16 \lesssim z \lesssim 20$, while the birth rate of Pop II
stars produces a much larger and broader peak at redshifts $2 \lesssim
z \lesssim 10$ (Ostriker \& Gnedin 1996; Gnedin \& Ostriker 1997;
Valageas \& Silk 1999).  Consequently, if GRBs are produced by the
collapse of massive stars in binaries, one expects them to occur out to
at least $z \approx 10$ and possibly $z \approx 15-20$, redshifts that
are far larger than those expected for the most distant quasars.  

Therefore, if GRBs -- or at least a well-defined subset of the
observed GRBs, such as the long bursts -- are due to the deaths of
massive stars, as theory and observations now suggest, then GRBs are a
powerful probe of the star-formation history of the universe, and
particularly of the SFR at VHRs.  In Figure 4, we have plotted the SFR
versus redshift from a phenomenological fit (Rowan-Robinson 1999) to
the star formation rate derived from submillimeter, infrared, and UV
data at redshifts $z < 5$, and from a numerical simulation by Gnedin \&
Ostriker (1997) (Figure 2b of their paper) at redshifts  $z \geq 5$. 
The simulations done by Gnedin \& Ostriker (1997) indicate that the SFR
increases with increasing redshift until $z \approx 10$, at which point
it levels off.  The smaller peak in the SFR at $z \approx 18$
corresponds to the formation of Population III stars, brought on by
cooling by molecular hydrogen.  In their other simulations, the
strength of this peak was found to be greater than in the example used
here (Ostriker \& Gnedin 1996; Gnedin \& Ostriker 1997).  Since GRBs
are detectable at these VHRs (see \S 2) and their redshifts may be
measurable from the absorption-line systems and the Ly$\alpha$ break in
the afterglows (see \S 5 below), if the GRB rate is proportional to the
star-formation rate, then GRBs could provide unique information about
the star-formation history of the VHR universe.

More easily but less informatively, one can examine the GRB peak photon
flux distribution $N_{GRB}(P)$.  To illustrate this, we have calculated
the expected GRB peak flux distribution assuming (1) that the GRB rate
is proportional to the star-formation rate\footnote{This may underestimate the GRB rate at VHRs since it is generally thought that the initial mass function will be tilted toward a greater fraction of massive stars at VHRs because of less efficient cooling due to the lower metallicity of the universe at these early times.}, (2) that the star-formation
rate is that given in Figure 4, and (3) that the peak photon luminosity
distribution $f(L_P)$ of the bursts is independent of $z$.  There is a
mis-match of about a factor of three between the $z < 5$ and $z \geq 5$
regimes.  However, estimates of the star formation rate are uncertain
by at least this amount in both regimes.  We have therefore chosen to
match the two regimes smoothly to one another, in order to avoid
creating a discontinuity in the GRB peak flux distribution that would
be entirely an artifact of this mis-match.

We calculate the observed GRB peak photon flux distribution
$N_{GRB}(P)$ as follows.  Assuming that GRBs are standard candles of
peak photon luminosity $L_P$, the peak photon flux distribution is
\begin{equation}
N_{GRB}(P|L_P) = \Delta T_{obs} 
\frac{R_{SF}(z)}{1+z}\frac{dV(z)}{dz}\left|\frac{dz(P|L_P)}{dP}\right|
\; ,
\end{equation}
where $\Delta T_{obs}$ is the length of time of observation, $R_{SF}(z)$ is the local co-moving star-formation rate at $z$, 
\begin{equation}
\frac{dV(z)}{dz} = 4\pi \frac{d_L^2(z)}{1+z}\left|\frac{dt(z)}{dz}\right|
\end{equation}
is the differential comoving volume,
\begin{equation}
d_L(z) = \cases{\frac{c}{H_0\sqrt{1-\Omega_m-\Omega_{\Lambda}}}
(1+z)\sinh{\left[\frac{H_0\sqrt{1-\Omega_m-\Omega_{\Lambda}}}{c}D(z)\right]} 
& ($\Omega_m+\Omega_{\Lambda} < 1$) \cr 
(1+z)D(z) & ($\Omega_m+\Omega_{\Lambda} = 1$) \cr 
\frac{c}{H_0\sqrt{\Omega_m+\Omega_{\Lambda}-1}}
(1+z)\sin{\left[\frac{H_0\sqrt{\Omega_m+\Omega_{\Lambda}-1}}{c}D(z)\right]} 
& ($\Omega_m+\Omega_{\Lambda} > 1$)}
\end{equation}
is the luminosity distance,
and 
\begin{equation}
\frac{dz(P|L_P)}{dP} = \left[\frac{dP(z|L_P)}{dz}\right]^{-1} \; .
\end{equation}
For $\Omega_m + \Omega_{\Lambda} = 1$,
\begin{equation}
\frac{dV(z)}{dz} = 4\pi D^2(z)\frac{dD(z)}{dz} \; ,
\end{equation}
where the comoving distance 
\begin{equation}
D(z) = \frac{c}{H_0}\int_0^z\frac{dz'}{[\Omega_m(1+z')^3 + 
\Omega_{\Lambda}]^{1/2}} \; ,
\end{equation}
and
\begin{equation}
\frac{dD(z)}{dz} = \frac{c}{H_0}\frac{1}{[\Omega_m(1+z)^3 +
\Omega_{\Lambda}]^{1/2}} \; .
\end{equation}
For $dL_P/d\nu \propto \nu^{-\alpha}$,
\begin{equation}
P(z|L_P) = \frac{L_P}{4\pi D^2(z)(1+z)^{\alpha}} \; .
\end{equation}
Again taking $\alpha = 1$, which is typical of GRBs (Mallozzi, Pendleton
\& Paciesas 1996), 
\begin{equation}
P(z|L_P) = \frac{L_P}{4\pi D^2(z)(1+z)} \; .
\end{equation}
Then 
\begin{equation}
\left|\frac{dP(z|L_P)}{dz}\right| = \frac{L_P}{4\pi}\left[\frac{2}{D^3(z)(1+z)}\frac{dD(z)}{dz} + \frac{1}{D^2(z)(1+z)^2}\right].
\end{equation} 
For a luminosity function $f(L_P)$ and for $dL_P/d\nu \propto 
\nu^{-\alpha}$, $N_{GRB}(P)$ is given by the following convolution integration:
\begin{equation}
N_{GRB}(P) = \Delta T_{obs} \int_0^{\infty}R_{GRB}(P|L_P)
f[L_P-4\pi D^2(z)(1+z)^{\alpha}P]dL_P \; .
\end{equation}

The upper panel of Figure 5 shows the number $N_*(z)$ of stars expected
as a function of redshift $z$ (i.e., the star-formation rate, weighted
by the co-moving volume, and time-dilated) for an assumed cosmology $\Omega_M = 0.3$ and
$\Omega_\Lambda = 0.7$ (other cosmologies give similar results).  The
solid curve corresponds to the star-formation rate in Figure 4.  The
dashed curve corresponds to the star-formation rate derived by Madau et
al. (1998).  This figure shows that $N_*(z)$ peaks sharply at $z
\approx 2$ and then drops off fairly rapidly at higher $z$, with a tail
that extends out to $z \approx 12$.  The rapid rise in $N_*(z)$ out to
$z \approx 2$ is due to the rapidly increasing volume of space.  The
rapid decline beyond $z \approx 2$ is due almost completely to the
``edge'' in the spatial distribution produced by the cosmology.  In
essence, the sharp peak in  $N_*(z)$ at $z \approx 2$ reflects the fact
that the  star-formation rate we have taken is fairly broad in $z$, and
consequently, the behavior of $N_*(z)$ is dominated by the behavior of
the co-moving volume $dV(z)/dz$; i.e., the shape of $N_*(z)$ is due
almost entirely to cosmology.  The lower panel in Figure 5 shows the
cumulative distribution $N_*(>z)$ of the number of stars expected as a
function of redshift $z$.  The solid and dashed curves have the same
meaning as in the upper panel.  This figure shows that $\approx 40\%$
of all stars have redshifts $z > 5$.

The upper panels of Figures 6, 7, and 8 show the predicted peak photon
flux distribution $N_{GRB}(P)$.  The solid curve assumes that all
bursts have a peak (isotropic) photon luminosity $L_P = 10^{58}$ ph
s$^{-1}$.  However, as remarked above, there is now overwhelming
evidence that GRBs are not ``standard candles.''  Consequently, we also
show in Figures 6 -- 8, as illustrative examples, the convolution of this
same star formation rate and a photon luminosity function, 
\begin{equation}
f(L_P) \propto \cases{L_P^{-\beta} & ($L_{\rm min} < L_P < L_{\rm max}$) \cr 
0 & (otherwise)}
\end{equation} 
where ($\log L_{\rm min}, \log L_{\rm max}$) = (57.5,58.5), (57,59),
and (56.5,59.5); i.e., $f(L_P)$ is centered on $L_P = 10^{58}$ ph
s$^{-1}$, and has widths $\Delta L_P / L_P = 10$, 100 and
1000.\footnote{The seven bursts with well-determined redshifts and
published peak (isotropic) photon luminosities have a mean peak photon
luminosity and sample variance $\log L_P = 58.1 \pm 0.7$.}  The actual
luminosity function of GRBs could well be even wider (Loredo \&
Wasserman 1998b, Lamb 1999). 

The general shape of the peak photon flux distributions $N_{GRB}(P)$
can be understood as follows.  The shape of $N_{GRB}(P)$ above the
rollover reflects the competition between the increasing number of GRBs
expected at larger $z$ (shown in Figure 5) and their decreasing peak
photon number flux $P(z)$ due to their increasing distance, while the
shape of $N_{GRB}(P)$ below the rollover reflects the intrinsic
luminosity function $f(L_P)$ of the bursts (Loredo \& Wasserman 1998). 
The latter is particularly the case because cosmology causes the
expected number of GRBs [$ \propto N_*(z)$] to have an ``edge,'' and
therefore to be sharply peaked in $z$, because of cosmology (Wasserman
1992).  Thus, most of the GRBs below the rollover in the peak photon
flux distribution $N_{GRB}(P)$ lie at the same distance ($z \approx 2$)
but have a range of intrinsic peak flux luminosities $L_P$, reflecting
the intrinsic luminosity function $f(L_P)$.  This is particularly clear
in Figure 6, where $N_{GRB}(P)$ is flat, and the plateau extends over
an  increasingly broad range of peak photon fluxes for increasingly
broad intrinsic  luminosity functions $f(L_P)$.  It is also evident in
Figures 7 and 8, where $N_{GRB}(P)$ below the rollover has slopes of
$\approx -1$ and  $\approx -2$.

Thus, information can be extracted from $N_{GRB}(P)$ about both the
GRB  rate as a function of redshift {\it and} the intrinsic luminosity 
function $f(L_P)$ of the bursts.  Figures 6, 7, and 8 show that the
limiting sensitivities of BATSE and HETE-2, and of {\it Swift} all lie well
below the observed rollover at $P \approx 6$ ph cm$^{-2}$ s$^{-1}$. 
Therefore, BATSE has detected, and HETE-2 will detect, many GRBs out
to $z \approx 10$, if this picture is correct.  {\it Swift} will
detect many GRBs out to $z \approx 14$, and will also detect for the
first time many intrinsically fainter GRBs.  

The middle panels of Figures 6 -- 8 show the predicted cumulative peak
photon flux distribution $N_{GRB}(> P)$ for the same set of luminosity
functions.  For the star formation rate that we have assumed, we find
that, if GRBs are assumed to be ``standard candles,'' the predicted
peak photon flux distribution falls steeply throughout the BATSE and
HETE-2 regime, and therefore fails to match the observed distribution,
in agreement with earlier work.  In fact, we find that a photon
luminosity function spanning at least a factor of 100 is required in
order to obtain semi-quantitative agreement with the principle features
of the observed distribution; i.e., a roll-over at a peak photon flux
of $P \approx 6$ ph cm$^{-2}$ s$^{-1}$ and a slope above this of about
-3/2.  This implies that there are large numbers of GRBs with peak
photon number fluxes below the detection threshold of BATSE and HETE-2,
and even of {\it Swift}.  

The lower panels of Figures 6 -- 8 show the predicted fraction of bursts with peak photon number flux $P$ that have redshifts of $z > 5$, for the same set of luminosity functions.  From these figures, we see that near the detection threshold of {\it Swift}, a significant number of bursts will have redshifts of $z > 5$.  Depending on the slope and the width of the luminosity function, more than half of such bursts may have redshifts of $z > 5$.


\section{GRBs as a Means of Finding Supernovae at Very High Redshifts}

GRBs can be used as beacons, revealing the locations of SNe at
high redshifts ($z > 1$).  To illustrate this, we take the best-fit
V-band light curve of the early afterglow of GRB 970228 from Reichart
(1999a) and add to it the V-band (or peak spectral flux) light curve of 
SN 1998bw (Galama et al. 1998; McKenzie \& Schaefer 1999), using the
light curve of SN 199bw as a template, as we did in Reichart (1999a). 
The light curves we use are corrected for Galactic extinction, as
explained in Reichart (1999a). We then transform the two components to
redshifts of $z = 1.2$, 3.0, 7.7, and 20, as described in \S 2.  
Figure 9 shows the resulting light curves.

Figure 9 shows that, if a SN 1998bw-like event occurred at a redshift
of $z \approx 3$, then it would peak in the K band about 70 days after
the event, and the peak magnitude would be K $\approx 24.4$. 
Consequently, the detection of high redshift SNe --- localized on the
sky and in redshift by earlier GRB afterglow observations --- is within
the limits of existing ground-based instruments, and well within the
limits of HST/NICMOS observations, out to a redshift of at least $z
\approx 3$.  At higher redshifts, SNe could be detected with NIR
observations at frequencies above the peak frequency of the SN in the
observer's frame; but because this portion of the SN spectrum is very
red, the flux at NIR frequencies in the observer's frame decreases
rapidly with increasing redshifts.  Consequently, SNe at redshifts
higher than $z \sim 4$ or 5 probably cannot be detected in the NIR
using existing instruments.  At still higher redshifts, one must appeal
to L- and M-band observations, but existing instruments do not yet have
the necessary sensitivity.

In Table 2, we expand upon Figure 9 by listing the band and the number
of days after the GRB that observations would have the best chance of
detecting a SN 1998bw-like event at peak flux density for a variety of
GRB redshifts.  We also list at what magnitudes SN 1998bw would have
been detected in these bands at these times, if transformed to these
redshifts.  Of course, the chances of detecting a SN component depend
on (1) how bright the afterglow is in the band of observation at the
time of observation, (2) how bright the host galaxy, if detected, is in
the band of observation, and (3) how much Galactic extinction there is
in the direction of the GRB in the band of observation.

Already one burst, GRB 971214, has been found to have a redshift, $z =
3.418$ (Kulkarni et al. 1998), that lies at the high end of the
redshift range for which current instruments can detect a SN
1998bw-like SN.  In the case of this burst, K-band observations were
taken 54 and 58 days after the burst (Ramaprakash et al. 1998), but
these observations did not go deep enough to detect a SN component
similar to SN 1998bw, were one present in the light curve.

As a further example, we consider the case of GRB 990510, a recent
burst whose afterglow faded as $t^{-2.4}$ (Stanek et al. 1999) or
$t^{-2.2}$ (Harrison et al. 1999) at late times; the difference in
these values can be traced to slight differences in these groups'
respective parameterizations of the light curve of this afterglow
(Harrison et al. 1999).  As a consequence of this rapid fading, a SN
1998bw-like component to the light curve, if present, would dominate
the afterglow after about a month at red and NIR wavelengths.  Again
using SN 1998bw as a template, we transform this template to the
redshift of the burst, $z = 1.619$ (Vreeswijk et al. 1999), and correct
this template for the difference in Galactic extinction  ($A_V = 0.233$
mag versus $A_V = 0.673$ mag) along the SN 1998bw and GRB 990510 lines
of sight, using the dust maps of Schlegel, Finkbeiner, \& Davis
(1998)\footnote{Software and data are available at
http://astro.berkeley.edu/davis/dust/index/html.}, and the Galactic
extinction curve of Cardelli, Clayton, \& Mathis (1989) for $R_V =
3.1$.  We plot the resulting observer-frame I-, J-, H-, and K-band
predictions for a SN 1998bw-like component of the afterglow of GRB
990510 at 49, 101, and 144 days after the burst in Figure 10.

If there is a SN component in the afterglow, it could have
easily been detected with the NICMOS instrument on HST, had it not run
out of cryogen  half a year earlier.  In the absence of NICMOS,
Fruchter et al. (1999b) performed HST/STIS observations of the
afterglow on 8.1 and 17.9 June 1999.  They detected the afterglow at V
$= 27.0 \pm 0.2$ mag and V $= 27.8 \pm 0.3$ mag on these two dates,
which is consistent with the extrapolated light curve of the afterglow,
and therefore is not consistent with an additional  SN 1998bw-like
component in the afterglow (Fruchter et al. 1999b).  However, caution
is in order, since this conclusion is subject to a number of
uncertainties. These include assumptions about (1) the spectral form of
the afterglow at the times of the observations, since the observations
spanned a wavelength range of 300 nm -- 900 nm, (2) how to extrapolate 
the light curve of the early afterglow of GRB 990510 to the times of
the observations; (3) the luminosity of the supernova component
relative to the luminosity of SN 1998bw, since Type Ib-Ic supernovae
are known not to be standard candles; (4) the brightness of SN 1998bw
at ultraviolet wavelengths; (5) the ignorability of any difference in
host galaxy extinction along the SN 1998bw and GRB 990510 lines of
sight; and (6) the underlying cosmological model.

\section{Measuring the Redshifts of Very High Redshift GRBs}

Of the eight GRBs with secure redshifts, four have redshifts between
$0.4 \la z \la 1$, two have redshifts between $1 \la z \la 2$, and one
(GRB 971214) has a redshift of $z = 3.42$ (see Table 1).  These
redshifts have been found in two ways:  (1) by taking a spectrum of the
afterglow at early times, when the afterglow was still sufficiently
bright (GRBs 970508, 980703, 990123, and 990510), and (2) by taking a
spectrum of the host galaxy, if detected, at sufficiently late times,
once the afterglow had faded (GRBs 970228, 970508, 971214, 980613, and
980703).  

Both methods have uncertainties.  In the former case, one technically
measures only a lower limit for the redshift of the burst,
corresponding to the redshift of the first absorber along the line of
sight from the burst.  However, as most, and possibly all, bursts with
optical afterglows are associated with host galaxies, this first
absorber is likely to be the host galaxy itself.  Consequently, GRB
redshifts measured in this way are fairly secure.  

In the latter case, one must establish that the positional coincidence
between the afterglow and the potential host galaxy is not accidental. 
For ground-based observations, $\approx$ 10\% of the sky  is covered by
galaxies brighter than R $\approx 25.5$, a typical magnitude for GRB
host galaxies (see, e.g., Hogg \& Fruchter 1999), due to seeing (Lamb
1999).  Consequently, identification of the host galaxy is best
established using HST images.  In the cases of GRB 970508 (Metzger et
al. 1997a, 1997b), GRB 980703 (Djorgovski et al. 1998), and GRB 990712
(Galama et al. 1999a), both absorption and emission lines were measured.
(It is notable that the redshifts for the two GRBs
at $z = 1.6$ were found by taking a spectrum of the GRB afterglow
at early times.  Currently, redshifts of host galaxies in the range $1
\la z \la 2.5$ are difficult to measure because the H$\alpha$ and [O
II] emission lines both lie outside the optical band for
this range of redshifts.)

At VHRs, e.g., $z \ga 5$, both methods will be challenging.  Consider
first the detection of absorption lines in afterglow spectra.   In
Figure 11, we plot the observed wavelengths of prominent absorption
lines as a function of redshift.  At VHRs, the prominent Balmer lines
are redshifted out of the NIR, and therefore out of the wavelength
range of instruments such as NIRSPEC (0.9 $\mu$m - 5.1 $\mu$m; McLean
et al. 1998).  Prominent metal lines such as Mg II and Fe II are not
redshifted out of the NIR.  However, both observations (see, e.g., York
1999) and theoretical calculations (see, e.g., Ostriker \& Gnedin 1996;
Gnedin \& Ostriker 1997; Valageas, Schaeffer, \& Silk 1999; Valageas \&
Silk 1999) suggest that the metallicity of the universe decreases with
increasing redshift, especially beyond $z \ga 3$.  Therefore, the
equivalent widths of the prominent metal lines are expected to decrease
with increasing redshift, making them challenging to detect at very
high redshifts.

However, prominent metal lines associated with the host galaxy may
still be present if many GRBs are due to the collapse of massive stars,
and the bursts occur near or in star-forming regions, since substantial
production of metals would be expected in the disk -- and certainly the
star-forming regions -- of the host galaxy.  This is illustrated by
Figure 11 of Valageas \& Silk (1999), which we have reprinted as Figure
12 in the present paper.  

The second method, which requires detecting the potential host galaxy, 
confirming its identification as the host galaxy through positional
coincidence with the GRB afterglow, and detecting Ly$\alpha$, [O II],
or other emission lines from the host galaxy, will also be challenging.
This will be the case because (1) galaxies more massive than $\sim
10^9$ $M_{\sun}$ are not expected to have formed by these times (see,
e.g., Ostriker \& Gnedin 1996; Gnedin \& Ostriker 1997), and (2)
because the surface brightness of these galaxies decreases as
$(1+z)^{-4}$.  However, at redshifts $z \ga 5$ (Fruchter
1999), and certainly at VHRs, the Ly$\alpha$  forest will be a very
prominent feature of the spectral flux distribution of the GRB
afterglow.  This is evident in Figure 2, which shows the expected flux
distribution of GRB afterglows at various redshifts.  It is even more
evident in Figure 13, which focuses in on NIR through optical
frequencies.

As an illustration of this, consider a burst that occurs at a redshift
slightly in excess of $z = 10$.  If its afterglow is similar to that of
GRB 970228, which had a relatively faint afterglow, its afterglow would be
detectable at K $\approx 16.2$ mag one hour after the burst, and at K
$\approx 21.6$ mag one day after the burst.  However, the afterglow
would not be detectable in the J band to any attainable limiting
magnitude.  Consequently, not only could VHR GRBs be detected and
identified relatively easily using existing ground-based instruments,
but given the extreme nature of this effect, accurate redshifts could
be determined from photometry alone.

One possible concern is that dust along the line of sight through the
star-forming region or the disk of the host galaxy could produce  
extinction (see, e.g., Reichart 1998, 1999b) that can mimic the
signature of the Ly$\alpha$ forest in the spectral flux distribution of
the afterglow (Lamb, Castander, \& Reichart 1999).  However, at VHRs
($z \ga 5$), this possibility is less likely due to (1) the lower
abundance of dust in the universe at these early times, and (2) the
increasing flux deficit of the Ly$\alpha$ forest with redshift.  We
illustrate this latter effect in the lower panel of Figure 13, in which
we re-plot the upper panel of Figure 13,  except that the solid curves
correspond to extincted versions of the solid curves.  In the lower
panel of Figure 13, we have taken $A_V = 1/3$ mag at the redshift of
the burst, an extinction magnitude that may be typical of the disks of
host galaxies (see, e.g., Reichart 1998), and we have adopted an
extinction curve that is typical of the interstellar medium of our
galaxy, using the extinction curve parameterization of Reichart
(1999b).  Figure 13 shows that, at redshifts higher than $z \approx 5$,
the signature of the Ly$\alpha$ forest clearly dominates the signature
of the extinction curve.

\section{Tracing the Metallicity of the Universe Using GRB Afterglows}


Recent studies of QSO absorption lines associated with damped
Ly$\alpha$ systems (Lu et al. 1996, Prochaska \& Wolfe 1997, Pettini
et al. 1997a,b) provide strong evidence that the metallicity of the
universe decreases with increasing redshift, and decreases dramatically
beyond $z \approx 3$.  Recent observations (Cowie et al. 1995) have
confirmed earlier evidence for a forest of C IV and Si IV doublets
associated with the forest of Ly$\alpha$ lines (Meyer \& York 1987). 
Observations of these systems extend to $z = 4.5$, higher than the
redshifts of the damped Ly$\alpha$ systems that have been observed to
date.  The detection of a forest of C IV and Si IV doublets, when 
combined with models of the ionization field from QSO radiation (see,
e.g., Meiksin \& Madau 1993, Cowie et al. 1995), suggests the existence
of a floor under the  abundances of heavy elements at roughly $10^{-2}$
of solar, extending out to the highest redshifts so far observed
(Songaila 1997).

These various abundance determinations indicate that heavy elements
exist in QSO absorption-line systems as early as $z = 5$,  although at
low levels, with a marked increase in the metallicity of these systems
evident at $z \approx 3$.  This metallicity history is consistent with
an early universal contamination of primordial gas by massive stars,
followed by a delay in forming additional heavy elements until $z
\approx 3$ (Timmes, Lauroesch \& Truran 1995), and finally a rise to
0.1 of solar abundances at $z = 2$.  The abundances of, e.g., Ca and Fe
inferred from QSO absorption-line systems  do not show a further
increase at $z < 1$ to fully solar values (Meyer \& York 1992).  This
may be due (1) to the fact that the disks of galaxies comparable to the
Milky Way, in which solar abundances exist, provide such a small cross
section for absorption against background quasars, compared to dwarf
galaxies (York 1999); (2) to the depletion of some heavy elements by
warm and cold clouds in low $z$ galaxies (Pettini et al. 1997a); and
(3) possibly to the fact that solar metal abundances may be anomalously
large by about a factor of three (Mushotsky 1999, private
communication).  It may be that all three of these factors play a role.

However, as we have seen, theoretical calculations of star formation 
in the universe predict that the earliest generation of stars occurs at
redshifts $z \approx 15 - 20$, and that the star formation rate
increases thereafter, peaking at $z \approx 2 - 10$ (Ostriker \& Gnedin
1996, Gnedin \& Ostriker 1997, Valageas \& Silk 1999).  One therefore
expects substantial metal production at $z > 3$.  The discrepancy
between this expectation and the abundances deduced from observations
of QSO absorption-line systems may reflect differences between the
metallicity of galactic disks and star-forming regions, and the
metallicities of the hydrogen clouds in the halos of galaxies and/or in
the IGM that are responsible for QSO absorption lines.

This possibility is supported by Figure 12, taken from Valageas \& Silk
(1999), which shows the redshift evolution of the metallicities of (1)
star-forming regions, (2) stars, (3) gas in galactic halos, and (4) the
overall average metallicity of matter given by their model
calculations.  Also shown are the data points from Pettini et al.
(1997b) for the metallicity of 34 damped Ly$\alpha$ systems, as
inferred from zinc absorption lines.  The curve corresponding to the
overall average metallicity of matter represents an upper bound for the
mean  metallicity of the IGM (corresponding to very efficient mixing). 
If galaxies do not eject metals very extensively into the IGM, the
metallicity of the IGM could be much smaller.  

Figure 12 shows that the mean metallicity expected for star-forming
regions is substantially more than that expected for clouds in the IGM
and for Ly$\alpha$ clouds associated with galactic halos (Lyman limit
or damped Ly$\alpha$ systems).  Consequently, it is possible that the
equivalent widths (EWs) of the absorption lines associated with the
host galaxy of the GRB (if it occurs in a galaxy) will remain large at
very high redshifts, even as the EWs of the absorption lines due either
to gas clouds in the IGM or associated with the halos of galaxies
weaken greatly beyond $z \approx 3$.  The situation at still higher
redshifts, where star formation may occur in globular cluster-sized
entities but not galaxies -- which have not yet had time to form -- is
unclear.

It is clear from this discussion that studies of absorption-line
systems in GRB afterglow spectra can contribute greatly to our
understanding of the metallicity history of the universe, and can allow
a comparison between the metallicity history of hydrogen clouds along
the line of sight to the burst {\it and} the metallicity history of the
star forming regions and/or disks of the burst host galaxies (and the
globular cluster-sized objects in which GRBs may occur at still higher
redshifts).

Core collapse SNe, such as the Type Ib-Ic SNe with which GRBs may be
associated, produce different relative abundances of various metals
than do Type Ia SNe, which are thought to be due to the thermonuclear
disruption of white dwarfs (see, e.g., Woosely \& Weaver 1986).  For
example, the spectrum of SN 1998bw, a peculiar Type Ic SN, exhibited
absorption lines reflecting the production of substantial amounts of O
and Cr, as well as Mg II, Fe, Ca, Si, and some S (Iwamoto et al. 1999,
Mazzoli et al. 2000).  This is typical of core collapse SNe (i.e., Type
Ib-Ic and Type II SNe).  Thus the relative abundances of various heavy
elements can also give clues about the origin of these elements; i.e.,
whether or not the metallicity is due primarily to Type Ib-Ic and Type
II SNe (core collapse SNe), and when and at what rate Type Ia SNe begin
to contribute to the increase in metallicity.  

Finally, studies of the absorption lines in GRB afterglow spectra can
help to determine whether the ratio [Fe/H] is a good chronometer at
high redshift, as has usually been assumed, or may not be, as recent
studies have suggested (Truran 1999, private communication).  Such
studies can also help to determine the extent and importance of mixing
between the metals and the hydrogen gas in galaxies and in the IGM.

\section{GRBs and Their Afterglows as Probes of Large-Scale Structure}

GRBs can be used to probe the large-scale structure of luminous matter
in the universe (Lamb and Quashnock 1993; Quashnock 1996).  The use of
GRBs  for this purpose has the advantage that they occur and are
detectable  out to VHRs, if GRBs are related to the collapse of massive
stars.  Thus observations of GRBs can be used to probe the properties
of large-scale structure at much higher redshifts (and much earlier)
than those that are currently probed by observations of galaxies and QSOs. 
GRBs also have the advantage that, while galaxy and QSO surveys on the
largest angular scales face difficulties due to the absorption of light
by dust and gas in the Galaxy, the Galaxy is completely transparent to
gamma rays.  Thus one can obtain a homogeneous sample of GRBs covering
the entire sky, unlike existing and future galaxy and QSO surveys.

On the other hand, the use of GRBs to probe large-scale structure
suffers from a very serious disadvantage:  Small number statistics. 
HETE-2 is expected to lead to the determination of the redshifts of a 
hundred or so bursts, while the {\it Swift} mission is likely
to lead to the determination of the redshifts of $\sim 1000$ bursts. 
These numbers are an order of magnitude smaller than the number of QSOs
whose redshifts are currently known, which is itself far smaller than
the  number of QSO redshifts expected from the Sloan Digital Sky
Survey.  Thus the use of GRBs themselves as a tracer of large-scale
structure in the universe may not be particularly powerful.

However, it should be possible to use the metal absorption lines and
the Ly$\alpha$ forest seen in the optical and infrared spectra of GRB
afterglows to probe the clustering of matter on the largest scales, as
has been done using these same lines in the optical spectra of QSOs
(Quashnock, Vanden Berk \& York 1996; Quashnock \& Vanden Berk 1998;
Quashnock \& Stein 1999).  Deep images of fields around QSOs with
absorbers in their spectra have revealed galaxies in the vicinities of
absorbers and at the same redshifts (see, e.g., Steidel, Dickenson \&
Persson 1994; Steidel et al. 1997).  A similar association has also
been inferred for a substantial fraction of the damped Ly$\alpha$
absorption lines (Lanzetta et al. 1995; Le Brun, Bergeron \& Boiss\'e
1996).  Consequently, it is thought that metal absorption lines are
associated with galaxy halos, and possibly galactic disks in some
cases.

As discussed in \S 2 and \S 3, we expect both GRBs and their afterglows
to occur and to be detectable out to very high redshifts ($z > 5$),
redshifts that are far larger than the redshifts expected for the most
distant quasars.  Consequently, the observation of absorption-line
systems and damped Ly$\alpha$ systems in the optical and infrared
spectra of GRB afterglows affords an opportunity to probe the
properties of these systems and their clustering at VHRs.   

At VHRs, one expects to be in the linear regime, and that the mass
spectrum of the Ly$\alpha$ forest systems, which are thought to lie in
the IGM, and possibly the damped Ly$\alpha$ systems, to follow the
Harrison-Zeldovich spectrum of density fluctuation in the early
universe.  Observations of NIR and infrared absorption lines in the
afterglow spectra of GRBs may allow one to test this expectation,
provided that these lines are detectable.  The decrease in metallicity
with increasing redshift may make it difficult to detect
absorption-line systems at redshifts $z \ga 5$, but this may not be
true for damped Ly$\alpha$ systems and the Ly$\alpha$ forest.

\section{GRB Afterglows as a Probe of the Epoch of Re-Ionization}

The epoch of re-ionization is one of the most important unknown
quantities relevant to large-scale structure and cosmology.  The lack
of Gunn-Peterson absorption implies that this epoch lies at $z > 5$,
while the lack of distortion of the microwave background by Compton
scattering off of free electrons implies that it lies at $z < 50$.  It
is plausible to have re-ionization occur anywhere in between these
redshifts in most cosmological models.  Observations of the afterglows
of VHR GRBs can be used to constrain the epoch of re-ionization by the
presence or absence of flux shortward of the Lyman limit in the rest
frame of the GRBs (see Figure 13). 

The absence of Gunn-Peterson troughs (Gunn \& Peterson 1965) in the
spectra of high-redshift quasars (Schneider, Schmidt, \& Gunn 1991) and
galaxies (Franx et al. 1997) indicates that the IGM was
re-ionized at a redshift in excess of $z \approx 5$.  Whether
re-ionization was caused by the first generation of stars, or by
quasars, is not yet known.  Assuming re-ionization was caused by the
first generation of stars, Gnedin \& Ostriker (1997) predict that
re-ionization occurred at $z \approx 7$; assuming re-ionization was
caused by quasars, Valageas \& Silk (1999) predict that re-ionization
occurred at $z \approx 6$.  However, Haiman \& Loeb (1998), assuming
that re-ionization was caused by stars and/or quasars, predict that
re-ionization occurred at a redshift in excess of $z \approx 11.5$.

Observations of VHR GRB afterglows may make it possible to distinguish
between these possibilities.   If re-ionization occurred at a redshift
of $z \la 6$, as Valageas \& Silk (1999) predict, then the redshift of
re-ionization could be measured directly from VHR GRB afterglow
photometry.  We show this in Figure 13, in which we re-plot the
spectral flux distributions shown in Figure 3, assuming that the
intergalactic medium was not re-ionized until a redshift of $4 < z < 5$
(dashed curves).  If re-ionization occurred at a redshift in excess of
$z \approx 6$, as Gnedin \& Ostriker (1997) and Haiman \& Loeb (1998)
predict, then observations of the afterglows of VHR GRBs could be used
to probe the redshift evolution of the Ly$\alpha$ forest out to this
redshift using existing ground-based instruments.  However, to reach $z
\approx 6$, observations would have to commence within a few hours of,
instead of $\approx 1$ day after, a burst.  The HETE-2 and
{\it Swift} missions, which will provide burst positions accurate to a
few arcseconds in near-real time, should make such observations
possible.  Future advances in instrumentation should allow redshifts in
excess of $z \approx 6$ to be probed.

\section{Conclusions}

The work of Bloom et al. (1999) in the case of GRB 980326, and the
subsequent work of Reichart (1999a) and Galama et al. (1999b) in the
case of GRB 970228, strongly suggest that at least some GRBs are
related to the supernovae of massive stars.  If  many GRBs are related
to the collapse of massive stars, we have shown that the bursts and
their afterglows can be used as a powerful probe of many aspects of the
very high redshift ($z \gtrsim 5$) universe.  

We have established that both GRBs and their afterglows are detectable
out to very high redshifts.  HETE-2 should detect GRBs out to $z
\approx 30$, while the {\it Swift} mission would be capable of
detecting GRBs out to $z \ga 70$, although it is unlikely that bursts
occur at such extremely high redshifts.

We have shown that, on the basis of theoretical calculations of star
formation in the universe, one expects GRBs to occur out to at least $z
\approx 10$ and possibly $z \approx 15-20$, redshifts that are far
larger than those expected for the most distant quasars.  This implies
that there are large numbers of GRBs with peak photon number fluxes
below the detection thresholds of BATSE and HETE-2, and even below the
detection threshold of {\it Swift}.  It also implies that HETE-2 will
detect many GRBs out to $z \approx 8$.  Similarly, the {\it
Swift} mission would detect many GRBs out to $z \approx 14$, and would
detect for the first time many intrinsically fainter GRBs.  The mere
detection of VHR GRBs would give us our first information about the
earliest generations of stars.  We have shown how GRBs and their
afterglows can be used as beacons to locate core collapse supernovae at
redshifts $z \gg 1$, and to study the properties of these supernovae.  

We have described the expected properties of the absorption-line
systems and the Ly$\alpha$ forest in the spectra of GRB afterglows. We
have described and compared various strategies for determining the 
redshifts of very high redshift GRBs.  We have shown how the
absorption-line systems and the Ly$\alpha$ forest visible in the
spectra of GRB afterglows can be used to trace the evolution of
metallicity in the universe, and to probe the large-scale structure of
the universe at very high redshifts.  Finally, we have shown how
measurement of the Ly$\alpha$ break in the spectra of GRB afterglows
can be used to constrain, or possibly measure, the epoch at which
re-ionization of the universe occurred, using the Gunn-Peterson test.

\acknowledgements

This research was supported in part by NASA grant NAG5-2868 and NASA
contract NASW-4690.  We thank Carlo Graziani for discussions about  the
proper way to calculate GRB peak photon flux and peak photon energy
distributions.  We thank Don York for discussions about QSO metal
absorption-line systems and the metallicity history of the universe,
and Jean Quashnock for discussions about using QSO absorption-line
systems to probe the large-scale structure of the universe.  We thank
Jim Truran for discussions about whether [Fe/H] is a good chronometer 
at early times in the evolution of the Galaxy and at high redshift.
Finally, we thank Dave Cole for providing information about the
limiting magnitudes that are detectable using current NIR instruments.

\clearpage

\begin{deluxetable}{ccccc} 
\tablecolumns{5}
\tablewidth{0pc} 
\tablecaption{Peak Photon Fluxes and Isotropic Luminosities for GRBs with
Secure Redshifts} \tablehead{\colhead{GRB} & \colhead{Redshift} &
\colhead{$P$ (ph cm$^{-2}$ s$^{-1}$)\tablenotemark{a}} & \colhead{$L_P$ (ph s$^{-1}$)\tablenotemark{b}} & \colhead{Redshift Reference}} 
\startdata
970228 & 0.695 & 3.5 & $5.1 \times 10^{57}$ & Djorgovski et al. 1999b \nl
970508 & 0.835 & 1.2 & $2.5 \times 10^{57}$ & Metzger et al. 1999a,b \nl
971214 & 3.418 & 2.3 & $6.4 \times 10^{58}$ & Kulkarni et al. 1998 \nl
980613 & 1.096 & 0.63 & $2.3 \times 10^{57}$ & Djorgovski et al. 1999a \nl
980703 & 0.967 & 2.6 & $7.4 \times 10^{57}$ & Djorgovski et al. 1998 \nl
990123 & 1.600 & 16.4 & $1.2 \times 10^{59}$ & Kulkarni et al. 1999 \nl
990510 & 1.619 & 8.16 & $6.2 \times 10^{58}$ & Vreeswijk et al. 1999 \nl
990712\tablenotemark{c} & 0.430 & -- & -- & Galama et al. 1999a \nl
\enddata \tablenotetext{a}
{From J. Norris (http://cossc.gsfc.nasa.gov/cossc/batse/counterparts). 
The listed peak photon flux is that in the energy band 50 - 300 keV.}
\tablenotetext{b}{Assuming $H_0 = 65$ km s$^{-1}$ Mpc$^{-1}$, 
$\Omega_m = 0.3$, and $\Omega_{\Lambda} = 0.7$.} 
\tablenotetext{c}{Peak photon flux not yet reported.} 
\end{deluxetable}

\clearpage

\begin{deluxetable}{ccccc}
\tablecolumns{5}
\tablewidth{0pc}
\tablecaption{Bands, Times, and Magnitudes at Which a SN 1998bw-Like Event 
Would Peak at Various Redshifts}
\tablehead{\colhead{Redshift} & \colhead{Band} & \colhead{Time (days)} & \colhead{Magnitude} & \colhead{Flux Density ($\mu$Jy)}}
\startdata
0.0 & V & 17 & -- & -- \nl
0.2 & R & 20 & 20.1 & 28 \nl
0.5 & I & 25 & 22.0 & 3.9 \nl
1.2 & J & 39 & 23.7 & 0.54 \nl
2.0 & H & 52 & 24.1 & 0.22 \nl
3.0 & K & 70 & 24.4 & 0.11 \nl
5.3 & L & 110 & 24.5 & 0.045 \nl
7.7 & M & 151 & 24.4 & 0.026 \nl
\enddata
\end{deluxetable}

\clearpage

\figcaption[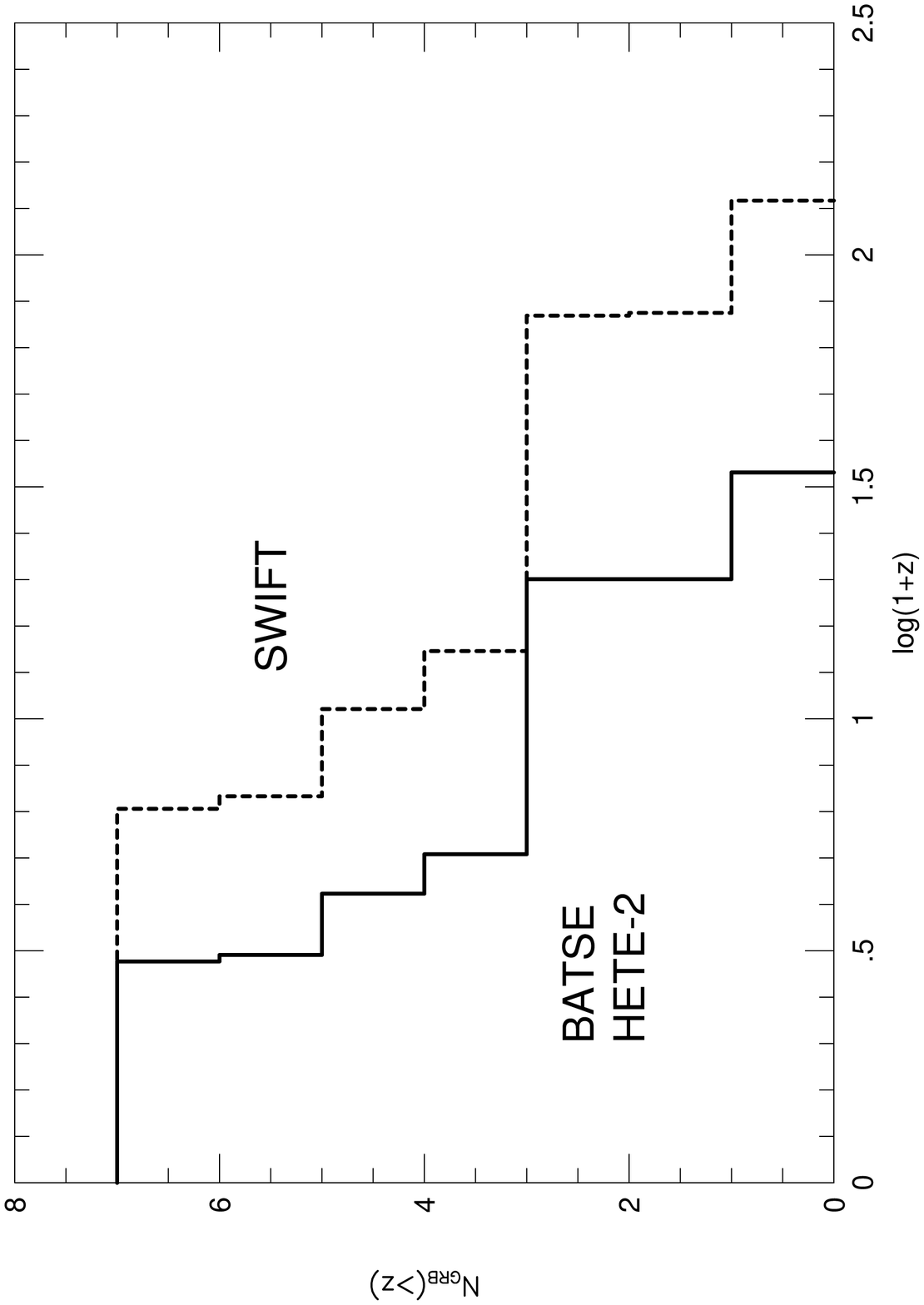] {Cumulative distributions of the limiting
redshifts at which the seven GRBs with well-determined redshifts and
published peak photon number fluxes would be detectable by BATSE and
HETE-2, and by {\it Swift}.\label{vhrfig2.ps}}

\figcaption[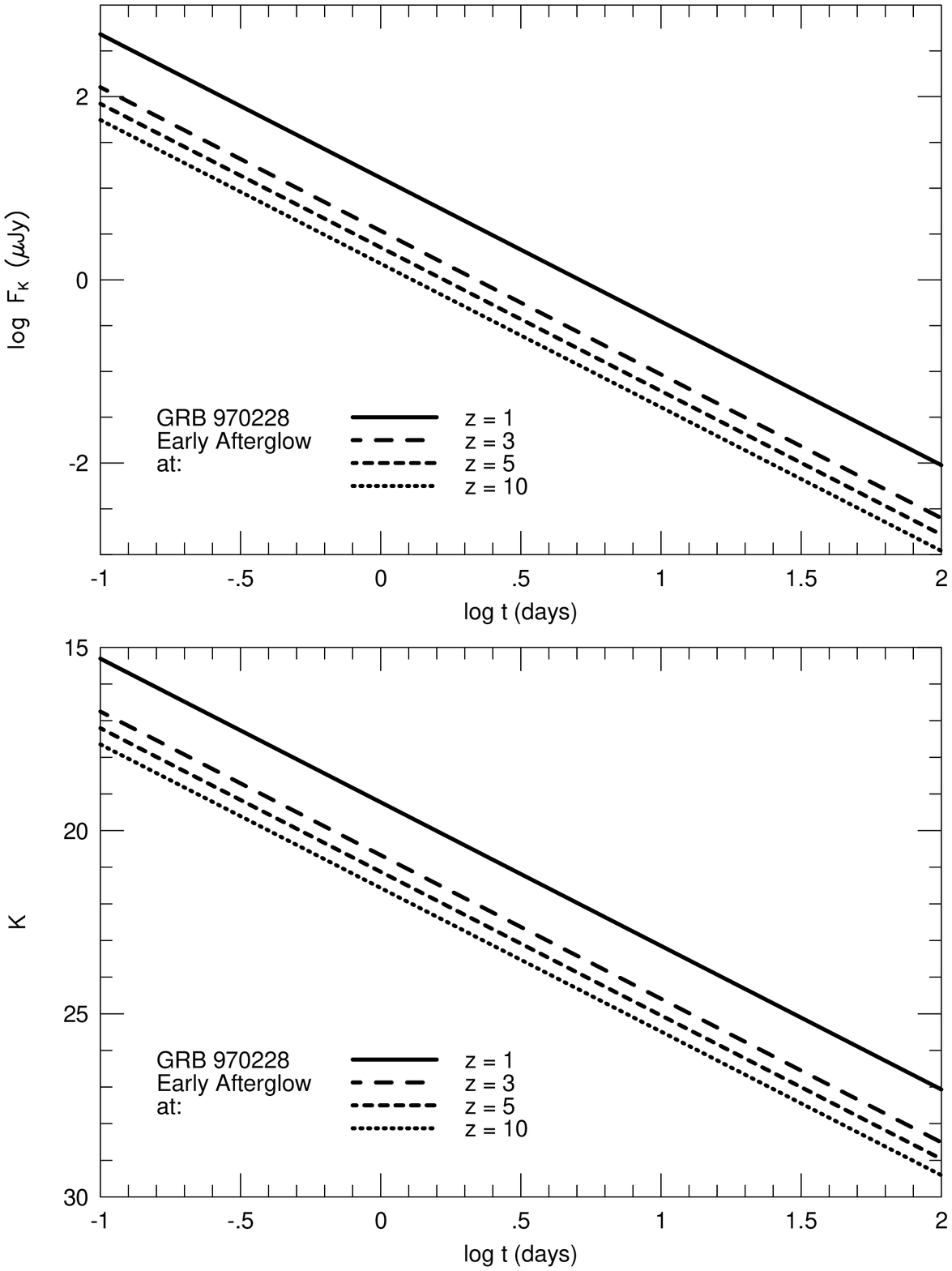]{The best-fit light curve of the early
afterglow of GRB 970228 from Reichart (1999a), transformed to various
redshifts.\label{vhrfig4.ps}} 

\figcaption[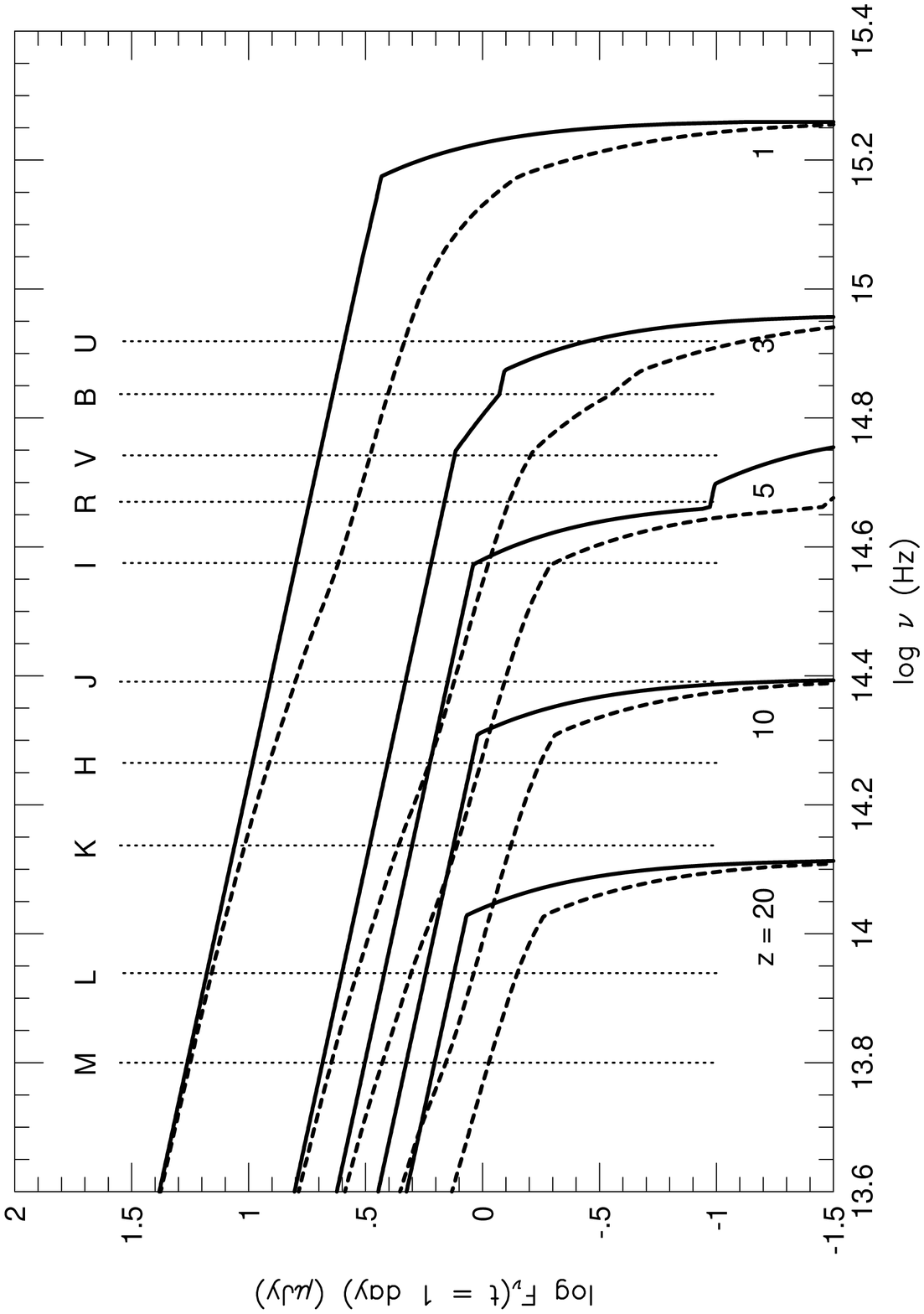]{The best-fit spectral flux distribution of the
early afterglow of GRB 970228 from Reichart (1999a), as observed one
day after the burst, after transforming it to various redshifts, and
extincting it with a model of the Ly$\alpha$ forest.  The dashed
curves are extincted versions of the solid curves, where we have
adopted $A_V = 1/3$ mag at the redshift of the burst, and an extinction
curve that is typical of the interstellar medium of our galaxy.\label{vhrfig3.ps}}

\figcaption[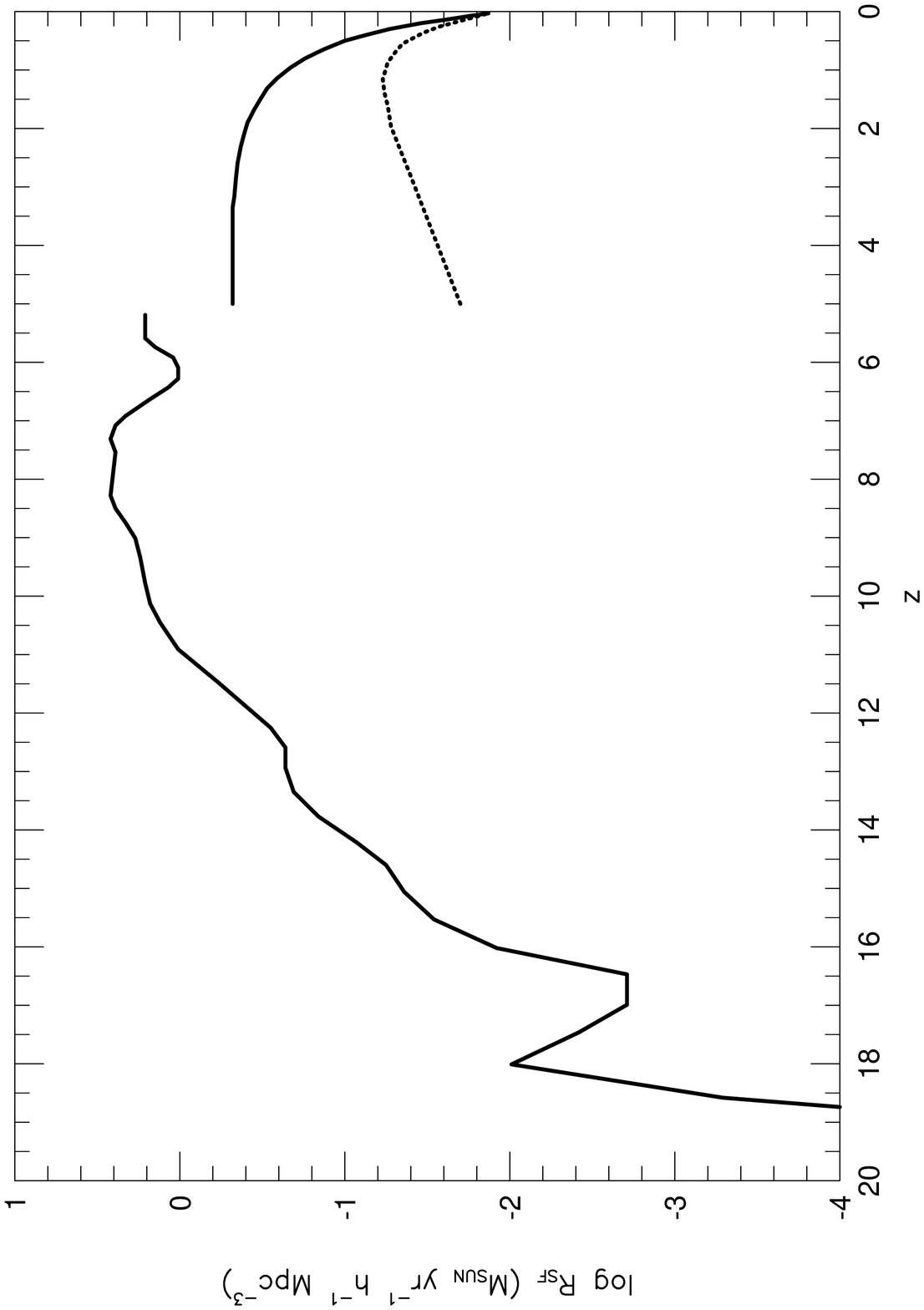]{The cosmic star-formation rate $R_{SF}$ as a
function of redshift $z$.  The solid curve at $z < 5$ is the star-formation rate derived by Rowan-Robinson (1999) from submillimeter,
infrared, and UV data; the solid curve at $z \ge 5$ is the star-formation rate calculated by Gnedin \& Ostriker (1997).  The dip in this curve at $z \approx 6$ is an artifact of their numerical simulation (Gnedin \& Ostriker 1997).  The dotted curve
is the star-formation rate derived by Madau et al. (1998).\label{vhrfig5.ps}}

\figcaption[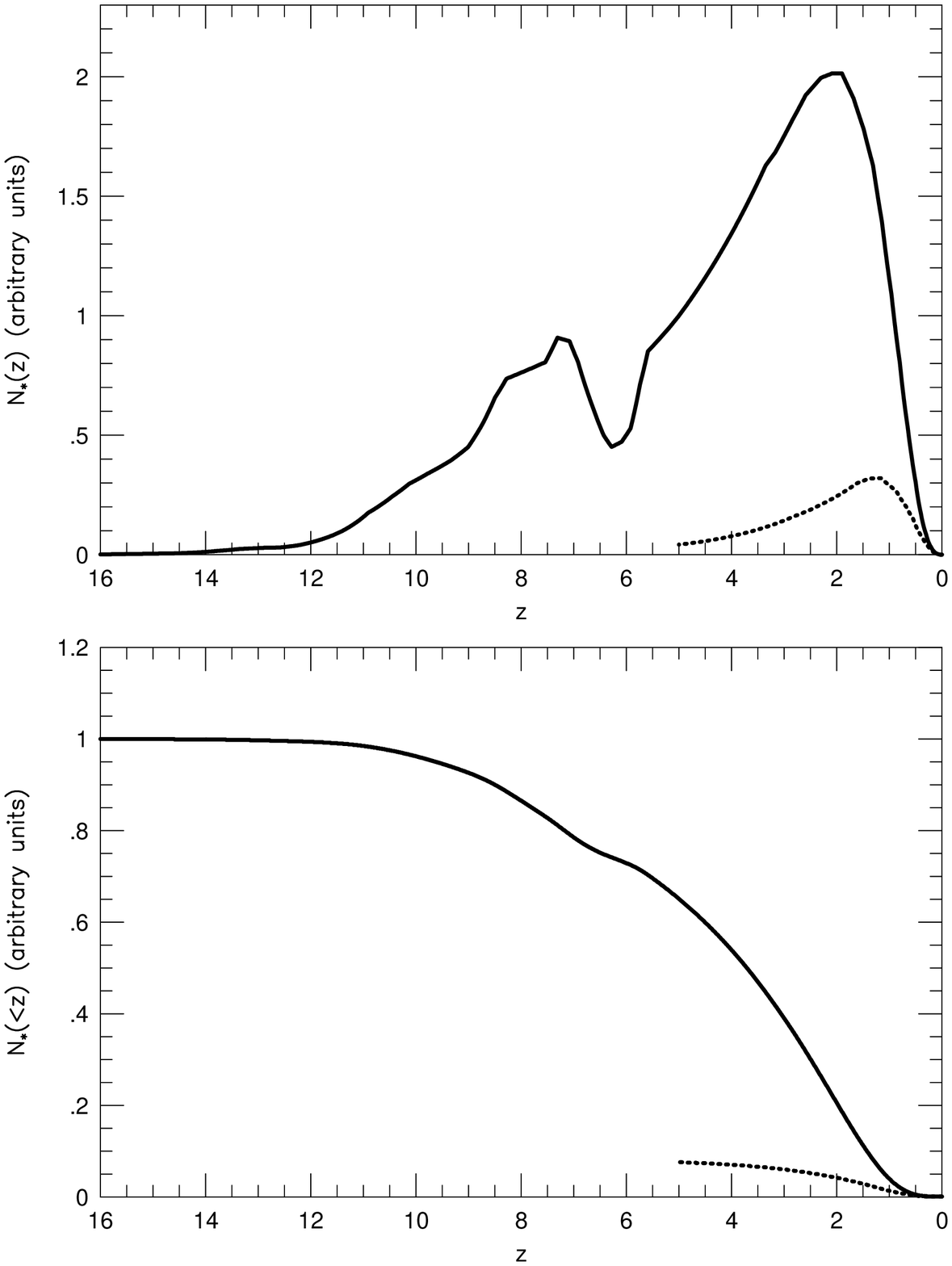]{Top panel:  The number $N_*$ of stars
expected as a function of redshift $z$ (i.e., the star-formation rate from Figure 4,
weighted by the differential comoving volume, and time-dilated) assuming that $\Omega_M = 0.3$
and $\Omega_\Lambda = 0.7$.  The solid and dashed curves, and the dip in the solid curve at $z \approx 6$, have the same
meanings as in Figure 4.  Bottom panel:  The cumulative
distribution of the number $N_*$ of stars expected as a function of
redshift $z$.  Again, the solid and dashed curves have the same meanings as in
Figure 4.  Note that $\approx 40\%$ of all stars have redshifts
$z > 5$.\label{vhrfig6.ps}}

\figcaption[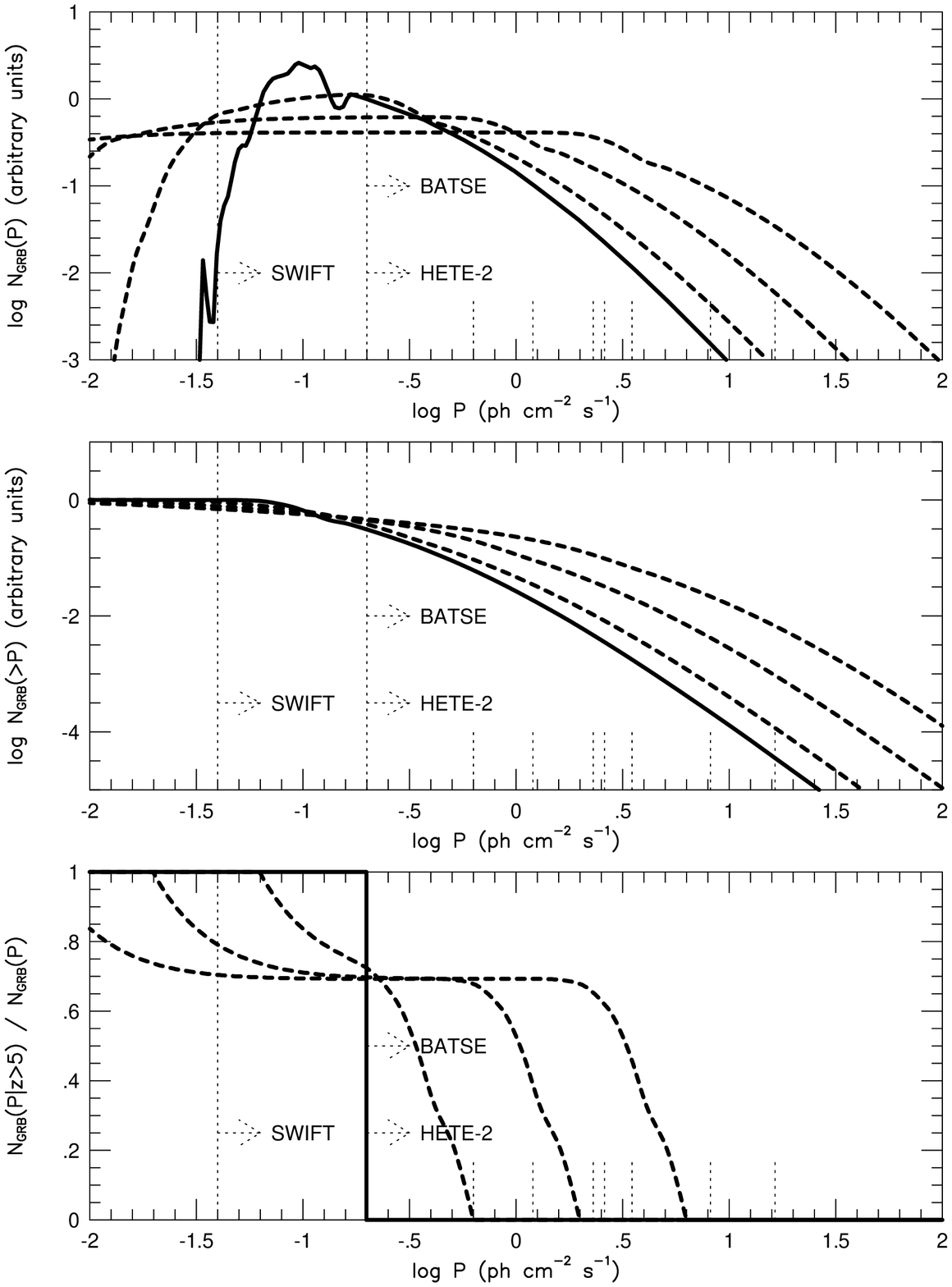]{Top panel:  The differential peak photon flux
distribution of GRBs, assuming that (1) the GRB rate is proportional to
the star-formation rate, (2) the star-formation rate is that shown in
Figure 5; and (3) the bursts are standard candles with a peak photon
luminosity $L_P = 10^{58}$ ph cm$^{-2}$ s$^{-1}$ (solid curve), or have
a logarithmically flat peak photon luminosity function that spans a
factor of 10, 100, or 1000 (dashed curves).  Approximate detection
thresholds are plotted for BATSE and HETE-2, and for {\it Swift}
(dotted lines).  The dotted hashes mark the peak photon fluxes of the
bursts from Table 1.  Middle panel:  The cumulative peak photon flux
distribution of GRBs for the same luminosity functions.  Lower panel: 
The fraction of GRBs with peak photon flux $P$ that have redshifts of
$z \ga 5$ for the same luminosity functions.\label{vhrfig7.ps}}

\figcaption[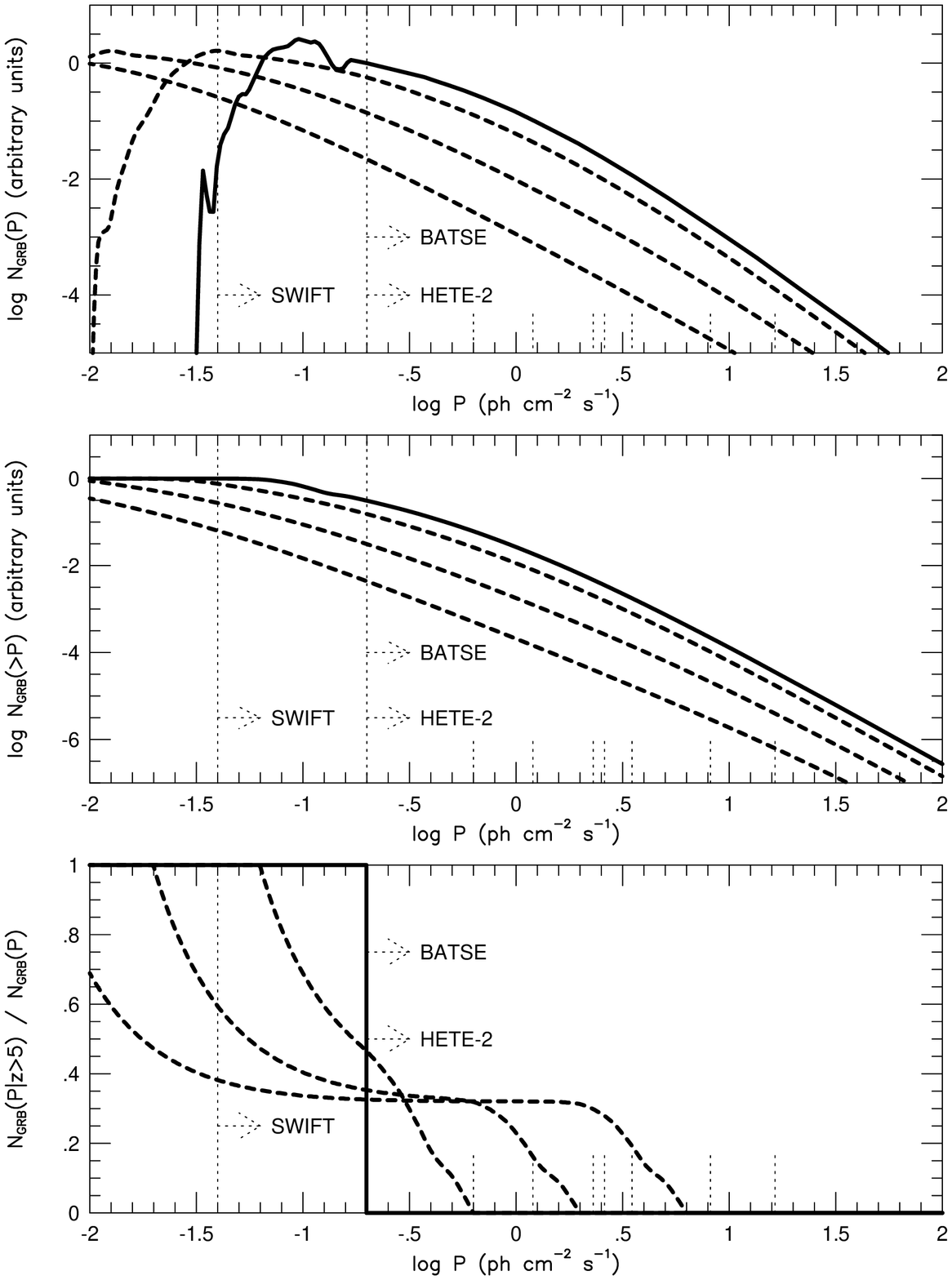]{Same as Figure 7, except that the dashed curves
correspond to a power-law GRB luminosity function with slope
$\beta = -1$.\label{vhrfig8.ps}}

\figcaption[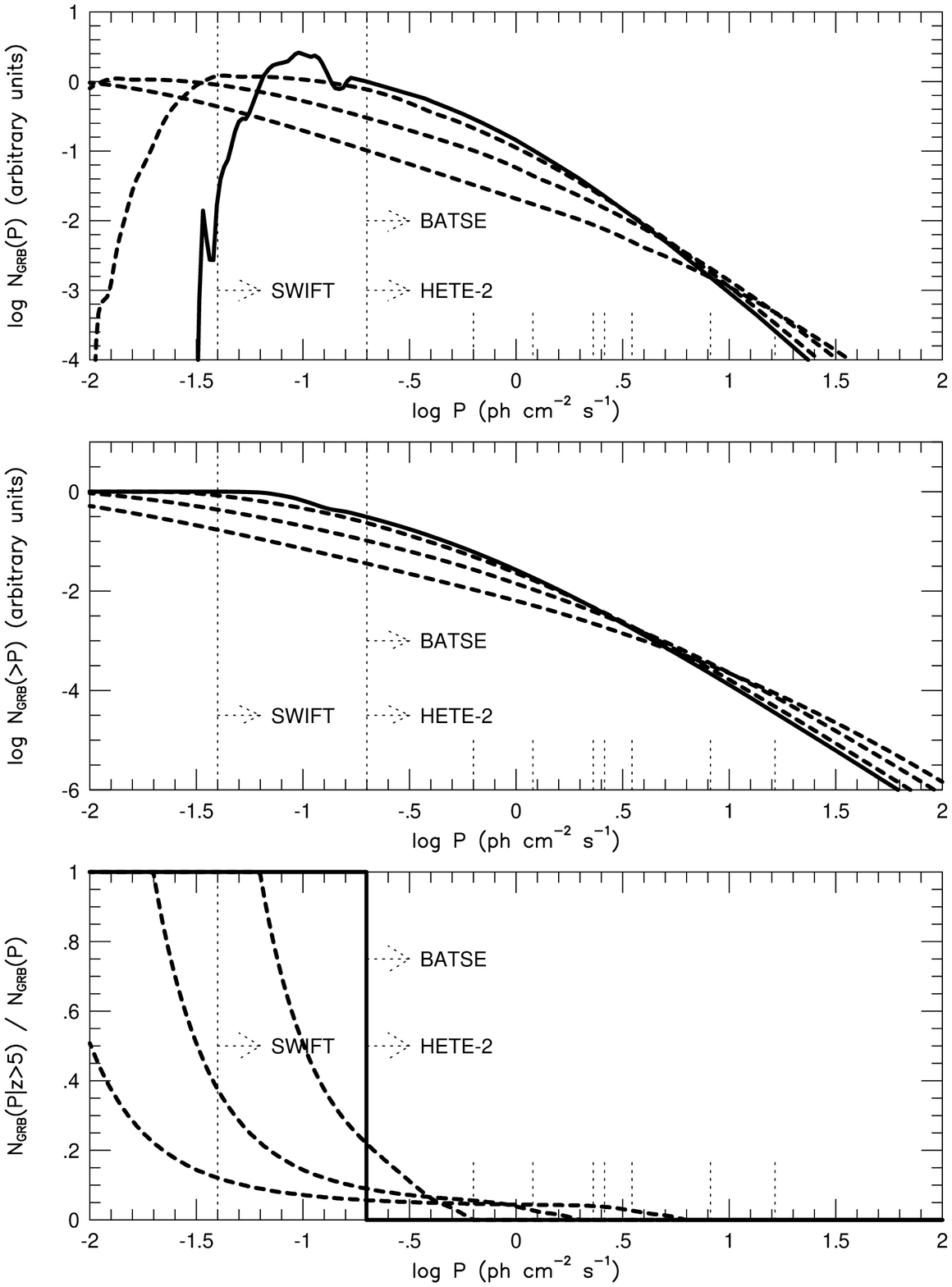]{Same as Figure 7, except that the dashed curves
correspond to a power-law GRB luminosity function with slope
$\beta = -2$.\label{vhrfig9.ps}}

\figcaption[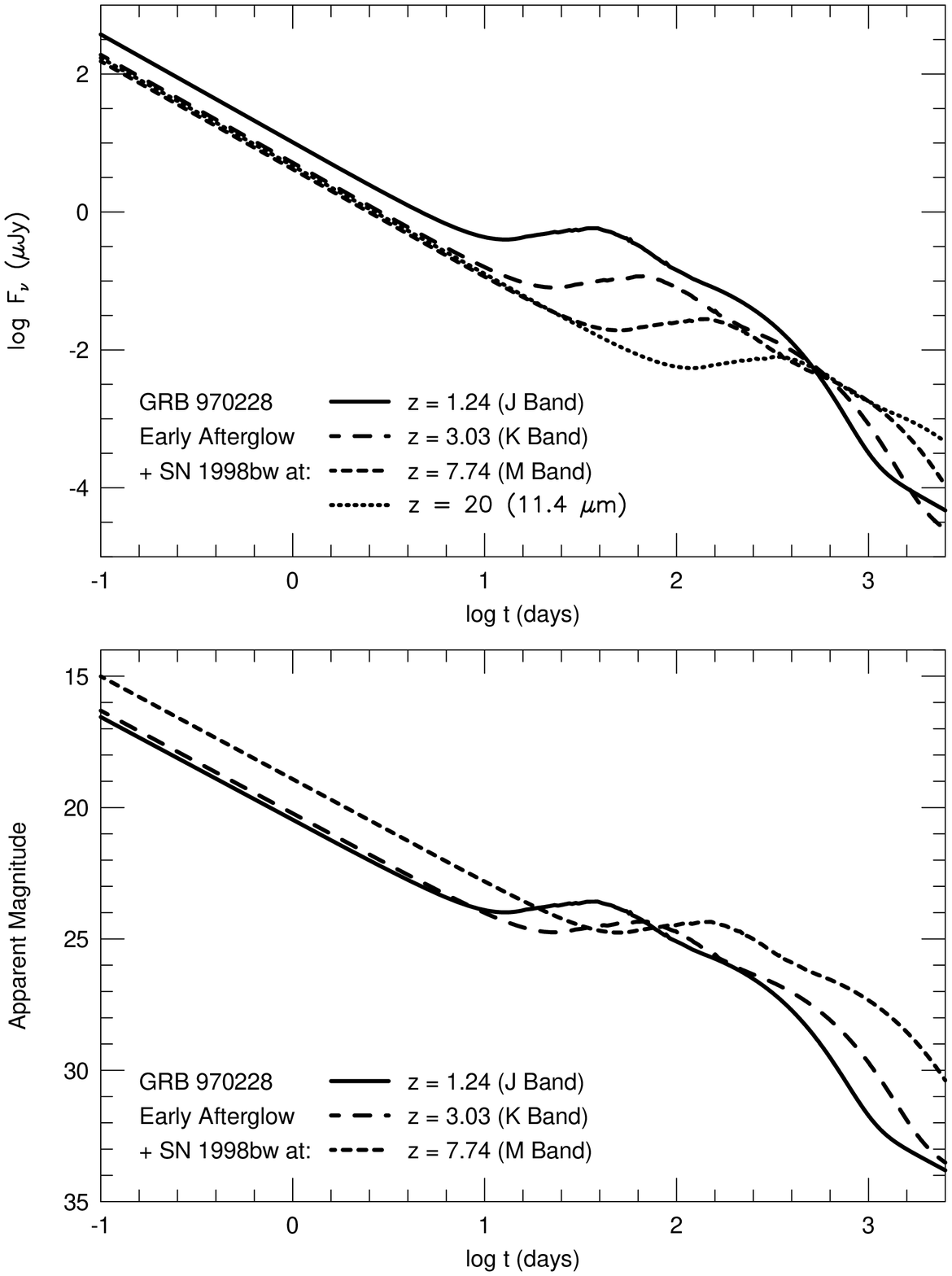]{The V-band, or peak flux density, light curve
of SN 1998bw plus the corresponding best-fit light curve of the early
afterglow of GRB 970228 from Reichart (1999a), transformed to various
redshifts, and corrected for Galactic extinction along the SN 1998bw
and GRB 970228 lines of sight (see \S 4 for more
details).\label{vhrfig10.ps}}

\figcaption[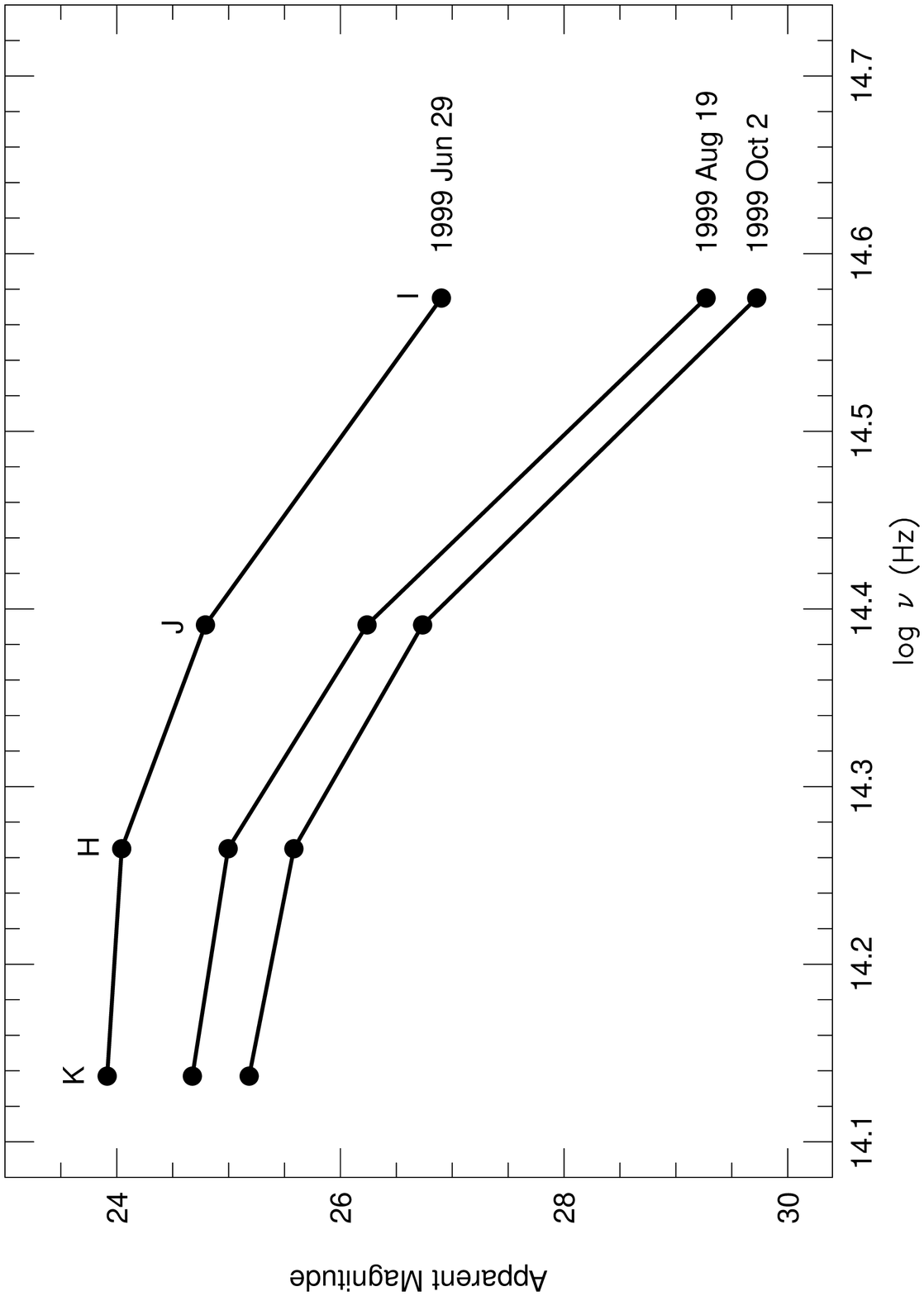]{Spectral flux distributions of SN 1998bw, as
observed 49, 101, and 144 days after the event, after transforming it
to the redshift, $z = 1.619$ (Vreeswijk et al. 1999), of GRB 990510,
and correcting it for differences in Galactic extinction along the SN
1998bw and GRB 990510 lines of sight.\label{vhrfig11.ps}}

\figcaption[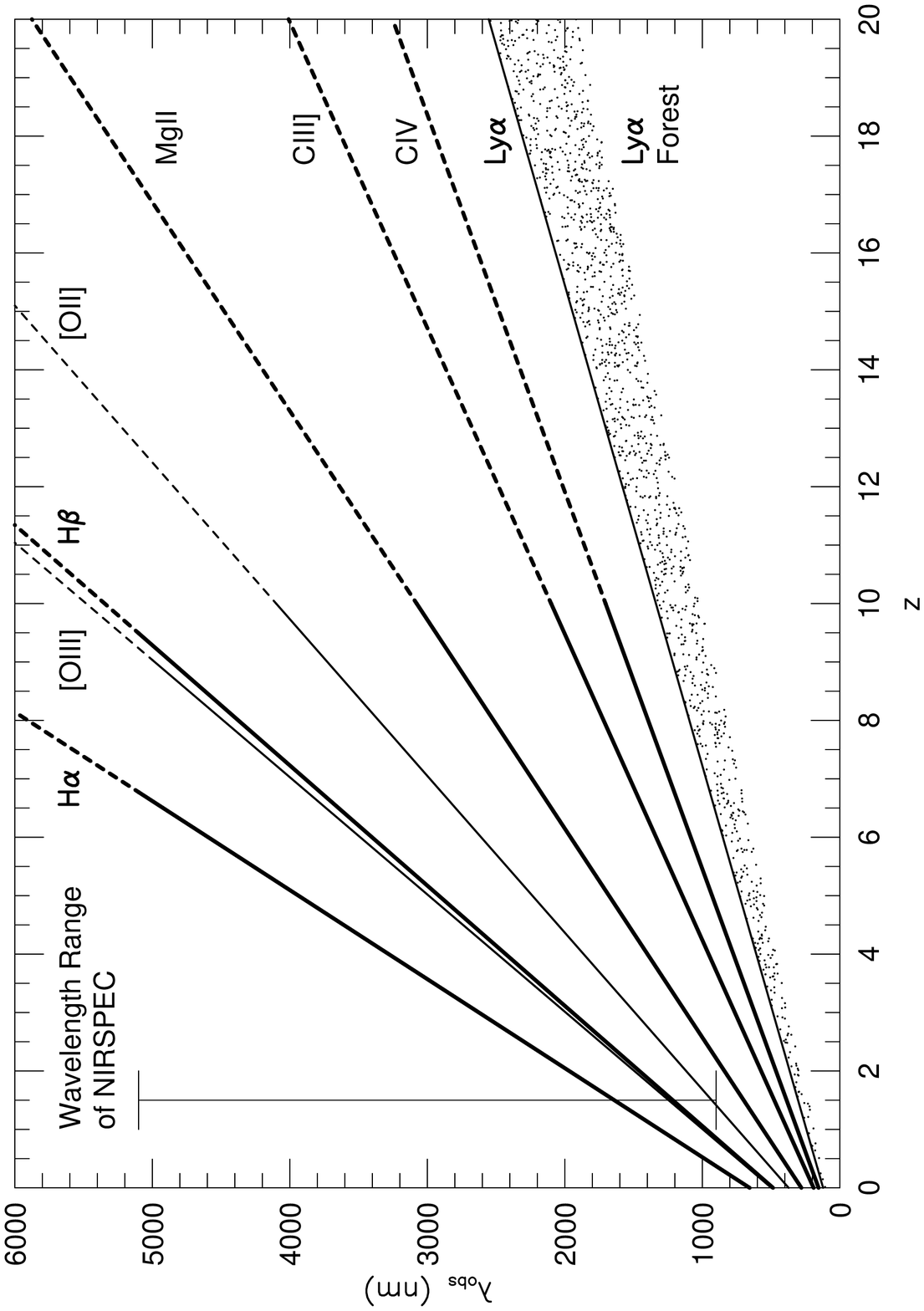]{The observed wavelengths of prominent
absorption lines and the Ly$\alpha$ forest as a function of redshift. 
At VHRs, the prominent Balmer lines will be difficult to detect because
they will be shifted out of the NIR, and the prominent metal lines may
be difficult to detect because of the low metallicity of the universe
at these early times (see Figure 12).\label{vhrfig12.ps}}

\figcaption[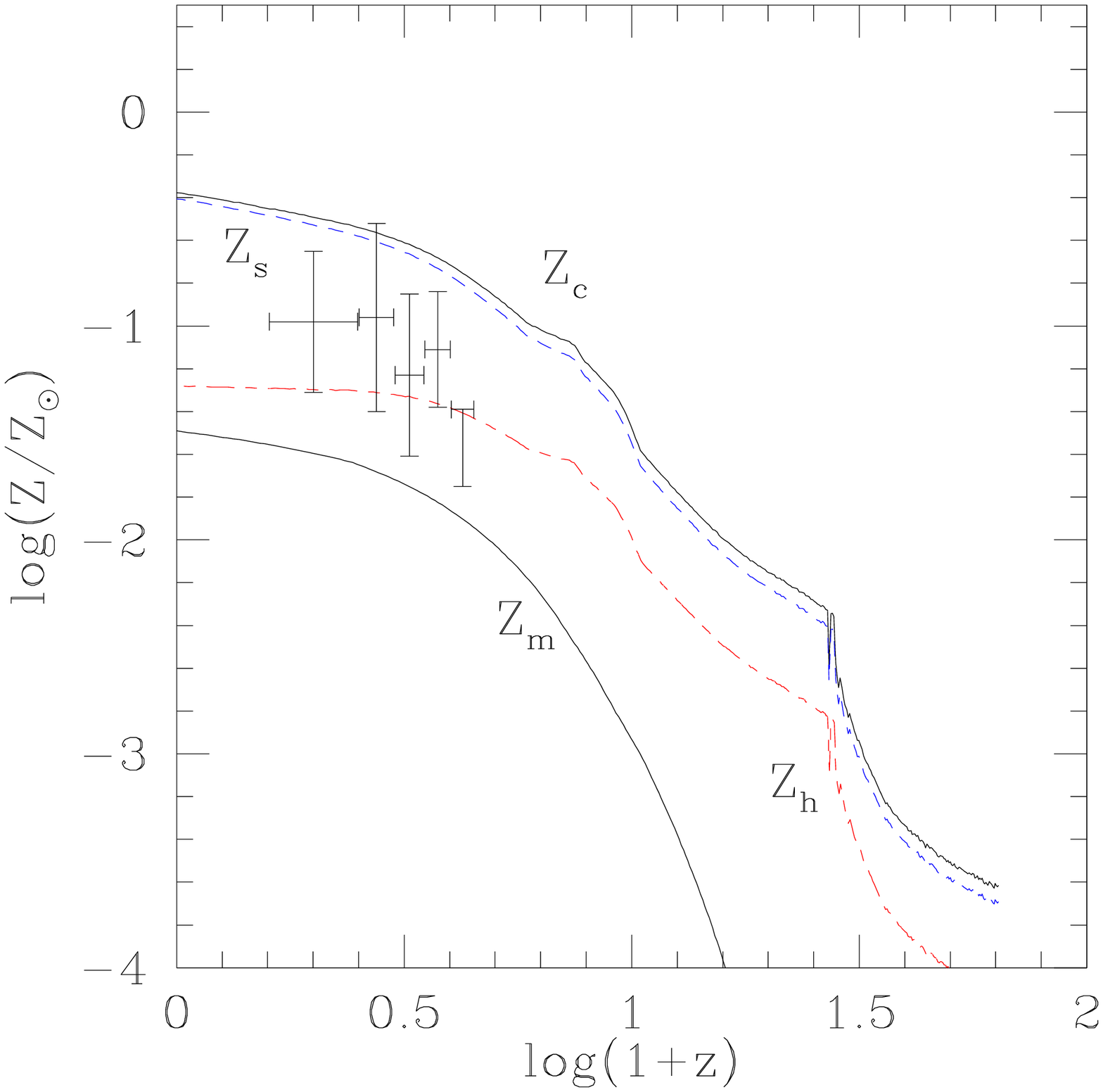]{Figure 9 of Valageas \& Silk (1999):  The
redshift evolution of the metallicities $Z_c$ (star-forming gas), $Z_s$
(stars), $Z_h$ (galactic halos), and $Z_m$ (matter average).  The data
points are from Pettini et al. (1997) for the zinc metallicity of
damped Ly$\alpha$ systems.\label{vhrfig13.ps}}

\figcaption[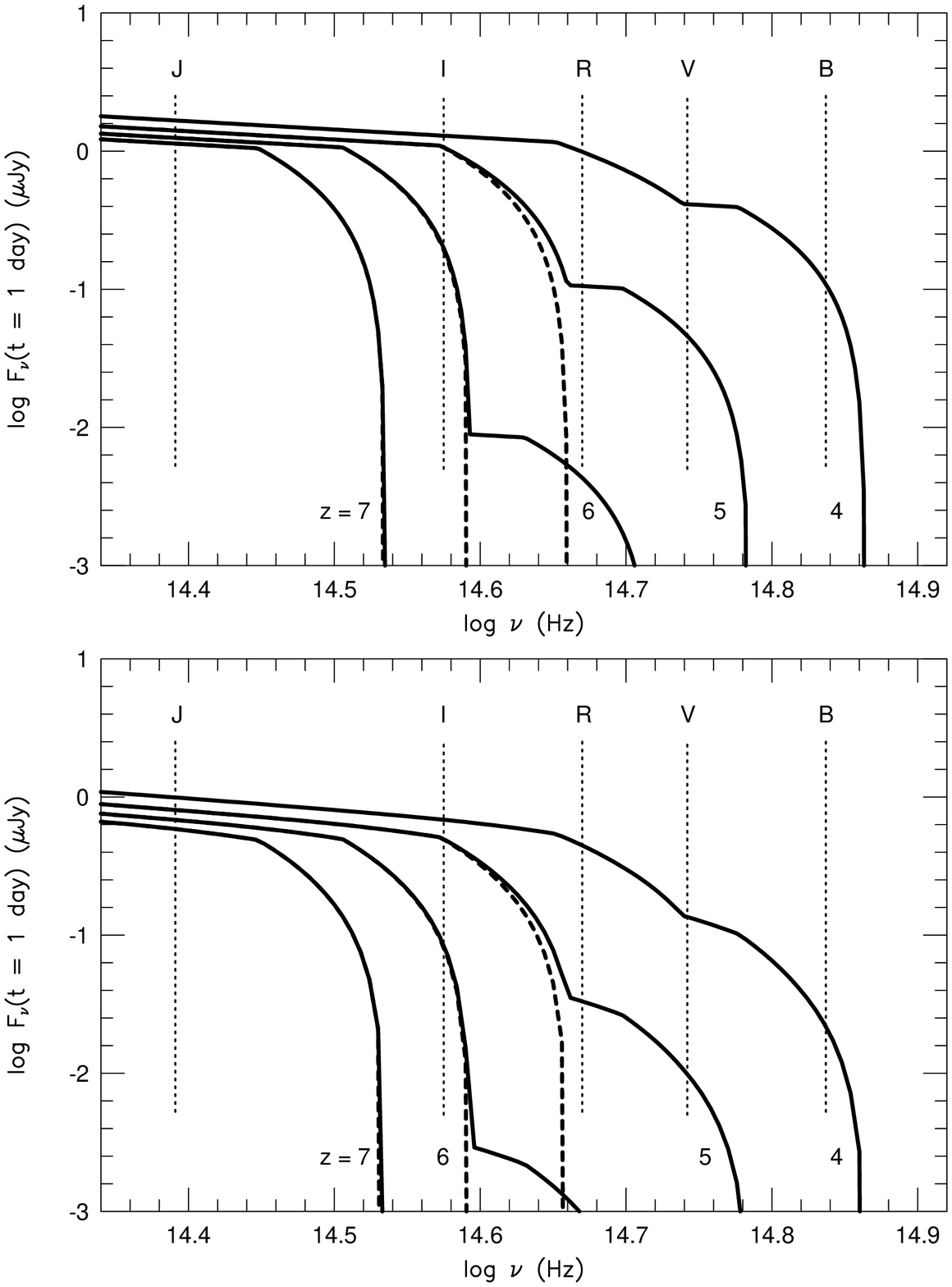]{Top Panel:  The solid curves are the same as
in Figure 3; we assume that reionization occurred at a redshift in
excess of $z = 7$.  The dashed curves are the same as the solid curves,
except that we instead assume that reionization occurred at a redshift
of $4 < z < 5$.  Bottom panel:  The solid curves are the same as in the
top panel, except extincted versions of the solid curves, where we
have adopted $A_V = 1/3$ mag at the redshift of the burst, and an
extinction curve that is typical of the interstellar medium of our
galaxy (see \S 7 for more details).  The dashed curves are the same as
the solid curves, except that we instead assume that reionization
occurred at a redshift of $4 < z < 5$.\label{vhrfig14.ps}}

\clearpage

\setcounter{figure}{0}

\begin{figure}[tb]
\plotone{vhrfig1.ps}
\end{figure}

\clearpage

\begin{figure}[tb]
\plotone{vhrfig3.ps}
\end{figure}

\clearpage

\begin{figure}[tb]
\plotone{vhrfig2.ps}
\end{figure}

\clearpage

\begin{figure}[tb]
\plotone{vhrfig4.ps}
\end{figure}

\clearpage

\begin{figure}[tb]
\plotone{vhrfig5.ps}
\end{figure}

\clearpage

\begin{figure}[tb]
\plotone{vhrfig6.ps}
\end{figure}

\clearpage

\begin{figure}[tb]
\plotone{vhrfig7.ps}
\end{figure}

\clearpage

\begin{figure}[tb]
\plotone{vhrfig8.ps}
\end{figure}

\clearpage

\begin{figure}[tb]
\plotone{vhrfig9.ps}
\end{figure}

\clearpage

\begin{figure}[tb]
\plotone{vhrfig10.ps}
\end{figure}

\clearpage

\begin{figure}[tb]
\plotone{vhrfig11.ps}
\end{figure}

\begin{figure}[tb]
\plotone{vhrfig12.ps}
\end{figure}

\clearpage

\begin{figure}[tb]
\plotone{vhrfig13.ps}
\end{figure}


\begin{thebibliography}{}

\bibitem[Bloom et al. 1998]{bea98}
Bloom, J.~S. et al. 1998, ApJ, 507, L25

\bibitem[Bloom et al. 1999]{bea99}
Bloom, J.~S., et al. 1999, Nature, in press (astro-ph/9905301)

\bibitem[Cardelli, Clayton, \& Mathis 1987]{ccd87}
Cardelli, J. A., Clayton, G. C., \& Mathis, J. S. 1987, ApJ, 345, 245

\bibitem[Castander \& Lamb 1998]{cl98}
Castander, F.~J. \& Lamb, D.~Q. 1998, in Gamma-Ray Bursts, eds. C.~A.
Meegan, R.~D. Preece, \& T.~M. Koshut (New York: AIP), 520

\bibitem[Castander \& Lamb 1999]{cl99}
Castander, F.~J. \& Lamb, D.~Q. 1999, ApJ, in press (astro-ph/9807195)

\bibitem[Connolly et al. 1997]{cea97}
Connolly, A.~J. 1997, ApJ, 486, L11

\bibitem[Costa et al. 1997]{cea97}
Costa, E., et al. 1997, IAU Circular 6572

\bibitem[Cowie et al. 1995]{cea95}
Cowie, G.~L., et al. 1995, AJ, 109, 1522

\bibitem[Djorgovski et al. 1998]{d98}
Djorgovski, S.~G., et al. 1998, ApJ, 508, L17

\bibitem[Djorgovski et al. 1999a]{d99a}
Djorgovski, S.~G., et al. 1999a, GCN Report 189

\bibitem[Djorgovski et al. 1999b]{d99b}
Djorgovski, S.~G., et al. 1999b, GCN Report 289

\bibitem[Franx et al. 1997]{fea97}
Franx, M., et al. 1997, ApJ, 486, L75

\bibitem[Frail \& Kulkarni 1997]{fk97}
Frail, D.~A., \& Kulkarni, S.~R. 1997, IAU Circular 6662

\bibitem[Fruchter 1999]{f99}
Fruchter, A. S. 1999, ApJ, 516, 683

\bibitem[Fruchter et al. 1999a]{fea99a}
Fruchter, A.~S., et al. 1999a, ApJ, in press (astro-ph/9807295)

\bibitem[Fruchter et al. 1999b]{fea99b}
Fruchter, A.~S., et al. 1999b, GCN Report 386

\bibitem[Galama et al. 1997]{gea97}
Galama, T.~J., et al. 1997, IAU Circular No. 6584

\bibitem[Galama et al. 1998]{gea98}
Galama, T.~J., et al. 1998, Nature, 395, 670

\bibitem[Galama et al. 1999a]{gea99a}
Galama, T.~J., et al. 1999a, GCN Report 388

\bibitem[Galama et al. 1999b]{gea99b}
Galama, T.~J., et al. 1999b, ApJ, submitted

\bibitem[Gallego et al. 1995]{gea95}
Gallego, J. 1995, ApJ, 455, L1

\bibitem[Gehrels 1999]{g99}
Gehrels, N. 1999, BAAS, 31, 993

\bibitem[Gnedin \& Ostriker 1997]{go97}
Gnedin, N.~Y., \& Ostriker, J.~P. 1997, ApJ, 486, 581

\bibitem[Gorosabel et al. 1998]{gea98}
Gorosabel, J., et al. 1998, A\&A, 335, L5

\bibitem[Graziani, Lamb, \& Marion 1999]{glm99}
Graziani, C., Lamb, D.~Q., \& Marion, G.~H. 1999, ApJ, in press

\bibitem[Gunn \& Peterson 1965]{gp65}
Gunn, J.~E., \& Peterson, B.~A. 1965, ApJ, 142, 1633

\bibitem[Haiman \& Loeb 1998]{hl98}
Haiman, Z., \& Loeb, A. 1998, ApJ, 503, 505

\bibitem[Halpern et al. 1999]{hea99}
Halpern, J.~P. 1999, ApJ, 517, L105

\bibitem[Harrison et al. 1999]{hea99}
Harrison, F.~A., et al. 1999, ApJ (Letters), in press

\bibitem[Hogg \& Fruchter 1999]{hf99}
Hogg, D.~W., \& Fruchter, A.~S. 1999, ApJ, 520, 54

\bibitem[Iwamoto et al. 1998]{iwa98}
Iwamoto et al. 1998, Nature, in press

\bibitem[Kouveliotou, et al. 1993]{kea93}
Kouveliotou, C., et al. 1993, ApJ, 413, L101

\bibitem[Kulkarni et al. 1998]{kea98}
Kulkarni, S.~R., et al. 1998, Nature, 395, 663

\bibitem[Kulkarni et al. 1999]{kea99}
Kulkarni, S.~R., et al. 1999, Nature, 398, 389

\bibitem[Lamb 1999]{l99}
Lamb, D.~Q. 1999, A\&A (Supplement), in press

\bibitem[Lamb, Castander, \& Reichart 1999]{lcr99}
Lamb, D.~Q., Castander, F.~J., \& Reichart, D.~E. 1999, A\&A (Supplement), 
in press

\bibitem[Lamb, Graziani, \& Smith 1993]{lgs93}
Lamb, D.~Q., Graziani, C. \& Smith, I.~A. 1993, ApJ, 413, L11

\bibitem[Lamb \& Quashnock 1993]{lq93}
Lamb, D.~Q., \& Quashnock, J.~M. 1993, in Gamma-Ray Bursts, eds. M. 
Friedlander, N. Gehrels, \& D. Macomb (New York: AIP), 1025

\bibitem[Lanzetta et al. 1995]{lea95}
Lanzetta, K.~M., et al. 1995, ApJ, 440, 435

\bibitem[Le Brun, Bergeron, \& Boiss\'e 1996]{lbb96}
Le Brun, V., Bergeron, J., \& Boiss\'e, P. 1996, A\&A, 306, 691

\bibitem[Lilly et al. 1996]{lea96}
Lilly, S.~J., et al. 1996, ApJ, 460, L1

\bibitem[Loredo \& Wasserman 1998]{lw98}
Loredo, T.~J., \& Wasserman, I.~M. 1998, ApJ, 502, 108

\bibitem[Lu et al. 1996]{lea96}
Lu, L., et al. 1996, ApJ (Supplement), 107, 475

\bibitem[Madau, Pozzetti, \& Dickinson 1998]{mpd98}
Madau, P., Pozzetti, L., \& Dickinson, M. 1998, ApJ, 498, 106

\bibitem[Mazzoli et al. 2000]{mas00}
Mazzoli, P. et al. 2000, ApJ, to be submitted

\bibitem[McKenzie \& Schaefer 1999]{ms99}
McKenzie, E.~H., \& Schaefer, B.~E. 1999, PASP, 111, 964

\bibitem[MacFadyen \& Woosley 1999]{mw99}
MacFadyen, A. I., \& Woosley, S. E. 1999, ApJ, 524, 262

\bibitem[MacFadyen, Woosley \& Heger 1999]{mw99}
MacFadyen, A. I., Woosley, S. E., \& Heger A. 1999, ApJ, submitted (astro-ph/9910034)

\bibitem[McLean et al. 1998]{mea98}
McLean, I.~S., et al. 1998, SPIE, 3354, 566

\bibitem[Meegan et al. 1993]{mea93}
Meegan, C.~A., et al. 1993, Second BATSE Catalog

\bibitem[Meyer \& York 1987]{my87}
Meyer, D.~M., \& York, D.~G. 1987, ApJ, 319, L45

\bibitem[Meyer \& York 1992]{my92}
Meyer, D.~M., \& York, D.~G. 1992, ApJ, 399, L121

\bibitem[M\'esz\'aros \& Rees 1993]{mes93}
M\'esz\'aros, P. \& Rees, M.~J. 1993, ApJ, 397, 570


\bibitem[Metzger et al. 1997a]{mea97a}
Metzger, M.~R., et al. 1997a, IAU Circular 6631

\bibitem[Metzger et al. 1997b]{metzger97b}
Metzger, M., et al. 1997b, Nature, 387, 878

\bibitem[Meiksin \& Madau 1993]{mm93}
Meiksin, A., \& Madau, P. 1993, ApJ, 412, 34

\bibitem[Narayan, Paczy\'nski, \& Piran]{npp92}
Narayan, R, Paczy\'nski, B., \& Piran, T. 1992, ApJ, 395, L83

\bibitem[Ostriker \& Gnedin 1996]{og96}
Ostriker, J.~P., \& Gnedin, N.~Y. 1996, ApJ, 472, L63

\bibitem[Paczy\'nski 1986]{p86}
Paczy\'nski, B. 1986, ApJ, 308, L43

\bibitem[Paczy\'nski 1998]{p98}
Paczy\'nski, B. 1998, ApJ, 494, L45

\bibitem[Pei, Fall & Hauser 1999]{pfh99}
Pei, Y. C., Fall, S. M., \& Hauser, M. G. 1999, ApJ, 522, 604

\bibitem[Pettini et al. 1997a]{pea97a}
Pettini, M., et al. 1997a, ApJ, 478, 536

\bibitem[Pettini et al. 1997b]{pea97b}
Pettini, M., et al. 1997b, ApJ, 486, 665

\bibitem[Prochaska \& Wolfe 1997]{pw97}
Prochaska, J.~X., \& Wolfe, A.~M. 1997, 474, 140

\bibitem[Quashnock 1996]{q96}
Quashnock, J.~M. 1996, ApJ, 461, L69

\bibitem[Quashnock \& Stein 1999]{qs99}
Quashnock, J.~M., \& Stein, M.~L. 1999, ApJ, 515, 506

\bibitem[Quashnock \& Vanden Berk 1998]{qv98}
Quashnock, J.~M., \& Vanden Berk, D.~E. 1998, ApJ, 500, 28

\bibitem[Quashnock, Vanden Berk, \& York 1996]{qvy96}
Quashnock, J.~M., Vanden Berk, D.~E., \& York, D.~G. 1996, ApJ, 472, L69

\bibitem[Ramaprakash et al. 1998]{rea98}
Ramaprakash, A.~N., et al. 1998, Nature, 393, 43

\bibitem[Reichart 1998]{r98a}
Reichart, D.~E. 1998, ApJ, 495, L99

\bibitem[Reichart 1999a]{r99a}
Reichart, D.~E. 1999a, ApJ, 521, L111

\bibitem[Reichart 1999b]{r99b}
Reichart, D.~E. 1999b, ApJ, in preparation

\bibitem[Ricker 1998]{r98}
Ricker, G. 1998, BAAS, 30, Abstract 33.14

\bibitem[Rowan-Robinson 1999]{r99}
Rowan-Robinson, M. 1999, Astroph. \& Space Sci., in press
(astroph/9906308)

\bibitem[Sahu \etal 1997a]{sea97a}
Sahu, K.~C., et al. 1997, Nature, 387, 476

\bibitem[Schlegel, Finkbeiner, \& Davis 1998]{sfd98}
Schlegel, D.~J., Finkbeiner, D.~P., \& Davis, M. 1998, ApJ, 500, 525

\bibitem[Schneider, Schmidt, \& Gunn 1991]{ssg91}
Schneider, D.~P., Schmidt, M., \& Gunn, J. E. 1991, AJ, 102, 837

\bibitem[Songaila 1997]{s97}
Songaila, A. 1997, ApJ, 490, L1

\bibitem[Stanek et al. 1999]{sea99}
Stanek, K.~Z., et al. 1999, ApJ (Letters), in press

\bibitem[Steidel et al. 1997]{sea97}
Steidel, C.~C., et al. 1997, ApJ, 480, 568

\bibitem[Steidel, Dickenson, \& Persson 1994]{sdp94}
Steidel, C.~C., Dickenson, M., \& Persson, S.~E. 1994, ApJ, 437, L75


\bibitem[Timmes, Lauroesch, \& Truran 1995]{tlt95}
Timmes, F.~X., Lauroesch, J.~T. and Truran, J.~W. 1995, ApJ, 451, 468

\bibitem[Totani 1997]{t97}
Totani, T. 1997, ApJ, 486, L71

\bibitem[Totani 1999]{t99}
Totani, T. 1999, ApJ, in press, (astro-ph/9805263)

\bibitem[Valageas, Schaeffer, \& Silk 1999]{vss99}
Valageas, P., Schaeffer, R., \& Silk, J. 1999, A\&A, 345, 691

\bibitem[Valageas \& Silk 1999]{vs99}
Valageas, P., \& Silk, J. 1999, A\&A, 347, 1

\bibitem[Vreeswijk et al. 1999]{vea99}
Vreeswijk, P. M., et al. 1999, GCN Report 324

\bibitem[Wasserman 1992]{was92}
Wasserman, I. 1992, ApJ, 394, 565

\bibitem[Wheeler et al. 1999]{wea99}
Wheeler, J. C., et al. 1999, ApJ, submitted (astro-ph/9909293)

\bibitem[Wijers et al. 1999]{wea99}
Wijers, R.~A.~M.~J., et al. 1999, MNRAS, 294, L13

\bibitem[Wijers, Rees, \& M\'esz\'aros 1997]{wrm97}
Wijers, R.~A.~M.~J., Rees, M.~J., \& M\'esz\'aros, P. 1997, MNRAS, 288,
L51

\bibitem[Woosley 1993]{w93}
Woosley, S.~E. 1993, ApJ, 405, 273

\bibitem[Woosley 1996]{w96}
Woosley, S.~E. 1996, in Gamma-Ray Bursts, eds. C.~A.
Meegan, R.~D. Preece, \& T.~M. Koshut (New York: AIP), 520

\bibitem[Woosely \& Weaver 1986]{ww86}
Woosley, S.~E., \& Weaver, T.~A. 1996, ARA\&A, 24, 205

\bibitem[York 1999]{y99}
York, D.~G. 1999, in Stromlo Workshop on High-Velocity Clouds, eds. B.~K. 
Gibson\& M.~E. Putman, ASP Conference Series, 166, 188

\bibitem[Zuo \& Lu 1993]{zl93}
Zuo, L., \& Lu, L. 1993, ApJ, 418, 601

\end{thebibliography}
\end{document}